\documentclass[aip,jcp,amsmath,amssymb,reprint]{revtex4-2}
\usepackage[utf8]{inputenc}
\usepackage[T1]{fontenc}
\usepackage[english]{babel}
\usepackage{graphicx}
\usepackage{lmodern}
\usepackage{color}
\usepackage{algorithm}
\usepackage{algpseudocode}
\usepackage[hidelinks]{hyperref}
\usepackage{mathptmx}
\usepackage{etoolbox}

\newcommand*{\E}{\mathop{}\!\mathrm{E}}

\newcommand*{\Var}{\mathop{}\!\mathrm{Var}}

\newcommand*{\diff}{\mathop{}\!\mathrm{d}}

\renewcommand*{\vec}[1]{\mathbf{#1}}

\renewcommand*{\tensor}[1]{\mathbf{#1}}

\newcommand\Pe{\mathrm{Pe}}

\graphicspath{{./figures/}}
\makeatletter
\def\input@path{{./figures/}}
\makeatother

\makeatletter
\def\@email#1#2{%
  \endgroup
  \patchcmd{\titleblock@produce}
  {\frontmatter@RRAPformat}
  {\frontmatter@RRAPformat{\produce@RRAP{*#1\href{mailto:#2}{#2}}}\frontmatter@RRAPformat}
  {}{}
}%
\makeatother
\begin{document}

\title[Adaptive Brownian Dynamics]{Adaptive Brownian Dynamics}

\author{Florian Sammüller}
\author{Matthias Schmidt}
\email{Matthias.Schmidt@uni-bayreuth.de}
\affiliation{Theoretische Physik II, Physikalisches Institut, Universität Bayreuth, D-95447 Bayreuth, Germany}

\date{\today}

\begin{abstract}
  A framework for performant Brownian Dynamics (BD) many-body simulations with adaptive timestepping is presented.
  Contrary to the Euler-Maruyama scheme in common non-adaptive BD, we employ an embedded Heun-Euler integrator for the propagation of the overdamped coupled Langevin equations of motion.
  This enables the derivation of a local error estimate and the formulation of criteria for the acceptance or rejection of trial steps and for the control of optimal stepsize.
  Introducing erroneous bias in the random forces is avoided by Rejection Sampling with Memory (RSwM) due to Rackauckas and Nie, which makes use of the Brownian bridge theorem and guarantees the correct generation of a specified random process even when rejecting trial steps.
  For test cases of Lennard-Jones fluids in bulk and in confinement, it is shown that adaptive BD solves performance and stability issues of conventional BD, already outperforming the latter even in standard situations.
  We expect this novel computational approach to BD to be especially helpful in long-time simulations of complex systems, e.g.\ in non-equilibrium, where concurrent slow and fast processes occur.
\end{abstract}

\maketitle

\section{Introduction}
Computer simulations have long become an established tool in the investigation of physical phenomena \cite{FrenkelUnderstandingMolecularSimulation2002,HansenTheorySimpleLiquids2013,AllenComputerSimulationLiquids2009}.
Complementing experimental results, they build the foundation for the exploration of increasingly complex dynamical systems.
From the standpoint of classical statistical mechanics, the simulation of a many-body system consisting of discrete interacting particles can reveal information about its structural correlation as well as its thermodynamic properties.
Naturally, this opens up the possibility of tackling many problems in the fields of material science, soft matter and biophysics, such as investigating the dynamics of macromolecules \cite{KarplusMolecularDynamicsSimulations2002}, predicting rheological properties of fluids \cite{SquiresSimpleParadigmActive2005, ReinhardtMicrorheologyCloseEquilibrium2014, PuertasMicrorheologyColloidalSystems2014}, or exploring non-equilibrium processes which occur e.g.\ in colloidal suspensions under the influence of external forcing \cite{GeigenfeindSuperadiabaticDemixingNonequilibrium2020}.

With the ever-increasing capabilities of computer hardware, a variety of different computational methods have emerged since the middle of the last century.
Conceptually, at least three distinct classes of particle-based simulation frameworks can be identified: i) Monte-Carlo (MC), which relies on the stochastic exploration of phase space, ii) Molecular Dynamics (MD), in which the set of ordinary differential equations (ODEs) of Hamiltonian dynamics is integrated to obtain particle trajectories, and iii) Langevin Dynamics, where random processes are incorporated into the Newtonian equations of motion so that the evolution of a system is obtained by numerical integration of then stochastic differential equations (SDEs).
Brownian Dynamics (BD) can be seen as a special case of iii), since the underlying stochastic Langevin equation is thereby considered in the overdamped limit where particle inertia vanishes and only particle coordinates remain as the sole microscopic degrees of freedom.

Notably, a broad range of refined methods have been developed in all three categories, sometimes even intersecting those.
Important examples of such extensions are kinetic Monte-Carlo for the approximation of time evolution from reaction rates \cite{GillespieGeneralMethodNumerically1976}, the addition of thermostats in MD to model thermodynamic coupling \cite{AndersenMolecularDynamicsSimulations1980,NoseMolecularDynamicsMethod1984}, event-driven algorithms which enable both MD and BD in hard-particle systems \cite{AlderStudiesMolecularDynamics1959,ScalaEventdrivenBrownianDynamics2007}, and the adaptation of molecular algorithms to modern hardware \cite{ColbergHighlyAcceleratedSimulations2011}.
Improvements in the calculation of observables from the resulting particle configurations have been made as well, e.g.\ by modifying their generation in MC (umbrella sampling, transition matrix MC \cite{WangTransitionMatrixMonte1999}, Wang-Landau sampling \cite{LandauNewApproachMonte2004}) or by utilizing advanced evaluation schemes in MD and BD, such as force sampling \cite{delasHerasBetterCountingDensity2018,RotenbergUseForceReduced2020,ColesReducedVarianceAnalysis2021} or adaptive resolution MD \cite{FritschAdaptiveResolutionMolecular2012}.

The efficiency and accuracy of a certain algorithm are always primary concerns, as these properties are essential for applicability and practicability in real-world problems.
One therefore aims to design procedures that are both as fast and as precise as possible -- yet it is no surprise that those two goals might often be conflicting.
Especially in BD, where stochastic processes complicate the numerical treatment, the development of more sophisticated algorithms apparently lacks behind that of MD, for example, and one often resorts to alternative or combined simulation techniques \cite{RosskyBrownianDynamicsSmart1978,JardatTransportCoefficientsElectrolyte1999}.
If the full dynamical description of BD is indeed considered, the equations of motion are usually integrated with the simple Euler-Maruyama method \cite{KloedenNumericalSolutionStochastic1992}, where stochasticity is accounted for in each equidistant step via normally distributed random numbers.
This can lead to inaccuracies and stability problems, making BD seem inferior to other computational methods.

In this work, we propose a novel approach to BD simulations which rectifies the above shortcomings of conventional BD.
To achieve this, we employ an adaptive timestepping algorithm that enables the control of the numerical error as follows.
The common Euler-Maruyama method is complemented with a higher-order Heun step to obtain an embedded integrator pair for an estimation of the local discretisation error per trial step.
By comparison of this error estimate with a prescribed tolerance, the trial step is either accepted or it is rejected and then retried with a smaller stepsize.
Particular care is required after rejections so as to not introduce a bias in the random forces which would violate their desired properties.
We therefore use Rejection Sampling with Memory (RSwM) \cite{RackauckasAdaptiveMethodsStochastic2017} to retain a Gaussian random process even in a scenario where already determined random increments may conditionally be discarded.
RSwM is a recently developed algorithm for the adaptive generation of random processes in the numerical solution of SDEs, which we improve and specialize to our context of overdamped Brownian motion and thereby formulate a method for adaptive BD simulations.

We demonstrate the practical advantages of adaptive BD over common BD in simulation results for prototypical bulk equilibrium systems and for more involved cases in non-equilibrium.
A notable example which we investigate is the drying of colloidal films at planar surfaces.
Especially when dealing with non-trivial mixtures, as e.g.\ present in common paints and coatings, the dynamics of this process can be inherently complex and its quantitative description turns out to be a major challenge \cite{vanderKooijWatchingPaintDry2015,SchulzCriticalQuantitativeReview2018}.
This stands in contrast to the necessity of understanding and predicting stratification processes in those systems.
Stratification leads to a dried film that has multiple layers differing in the concentration of constituent particle species, thereby influencing macroscopic properties of the resulting colloidal film.
Therefore, controlling this process is an important measure to tailor colloidal suspensions to their field of application.
Advances in this area have been made experimentally \cite{MakepeaceStratificationBinaryColloidal2017,TruemanAutostratificationDryingColloidal2012}, by utilizing functional many-body frameworks like dynamical density functional theory (DDFT) \cite{HeDynamicalDensityFunctional2021}, and with molecular simulations such as conventional BD \cite{FortiniStratificationSizeSegregation2017}.
By employing the adaptive BD method, we are able to capture the complex dynamical processes occuring in those systems even in the final dense state.
Close particle collisions and jammed states are resolved with the required adjustment of the timestep, necessary for the stability and accuracy of the simulation run in those circumstances.
This can not be achieved easily with common BD.

This paper is structured as follows.
In Sec.\ \ref{sec:Numerics_SDEs}, a brief and non-rigorous mathematical introduction to the numerical solution of SDEs is given.
Particularly, we illustrate the prerequisites for adaptive and higher-order algorithms in the case of general SDEs and emphasize certain pitfalls.
In Sec.\ \ref{sec:Application_to_BD}, these considerations are applied to the case of Brownian motion.
We construct the embedded Heun-Euler integration scheme in Sec.\ \ref{subsec:Embedded_Heun-Euler} and incorporate RSwM in Sec.\ \ref{subsec:RSwM_BD}, which yields the adaptive BD method.
Observables can then be sampled from the resulting particle trajectories with the means illustrated in Sec.\ \ref{subsec:Sampling_observables}.
In Sec.\ \ref{sec:Simulation_results}, simulation results of the above mentioned Lennard-Jones systems are shown and the practical use of adaptive BD is confirmed.
In Sec.\ \ref{sec:Conclusion}, we conclude with a summary of the shown concepts, propose possible improvements for the adaptation of timesteps, and present further ideas for use cases.

\section{Numerics of Stochastic Differential Equations}
\label{sec:Numerics_SDEs}
Brownian Dynamics of a classical many-body system of $N$ particles in $d$ spatial dimensions with positions $\vec{r}^{N}(t) = (\vec{r}^{(1)}(t), \dots, \vec{r}^{(N)}(t))$ at time $t$ and temperature $T$ is described by the overdamped Langevin equation.
The trajectory of particle $i$ satisfies
\begin{equation}
  \label{eq:overdamped_Langevin}
  \dot{\vec{r}}^{(i)}(t) = \frac{1}{\gamma^{(i)}} \vec{F}^{(i)}(\vec{r}^N(t)) + \sqrt{\frac{2 k_B T}{\gamma^{(i)}}} \vec{R}^{(i)}(t)
\end{equation}
where $\vec{F}^{(i)}(\vec{r}^N(t))$ is the total force (composed of external and interparticle contributions) acting on particle $i$, $\gamma^{(i)}$ is the friction coefficient of particle $i$ and $k_B$ is Boltzmann's constant; the dot denotes a time derivative.
In eq.\ \eqref{eq:overdamped_Langevin}, the right-hand side consists of a deterministic (first summand) and a random contribution (second summand).
The random forces are modeled via multivariate Gaussian white noise processes $\vec{R}^{(i)}(t)$ which satisfy
\begin{align}
  \langle \vec{R}^{(i)}(t) \rangle &= 0,\\
  \langle \vec{R}^{(i)}(t) \vec{R}^{(j)}(t') \rangle &= \tensor{I} \delta_{ij} \delta(t - t'),
\end{align}
where $\langle \cdot \rangle$ denotes an average over realizations of the random process, $\tensor{I}$ is the $d \times d$ unit matrix, $\delta_{ij}$ denotes the Kronecker delta, and $\delta(\cdot)$ is the Dirac delta function.

One can recognize that eq.\ \eqref{eq:overdamped_Langevin} has the typical form of an SDE
\begin{equation}
  \label{eq:SDE_general}
  \diff X(t) = f(X(t), t) \diff t + g(X(t), t) \diff W(t),
\end{equation}
if the dependent random variable $X$ is identified with the particle positions $\vec{r}^N$ and $W$ is a Wiener process corresponding to the integral of the Gaussian processes $\vec{R}^N = (\vec{R}^{(1)}, \dots, \vec{R}^{(N)})$.
As we do not consider hydrodynamic interactions, the random forces in eq.\ \eqref{eq:overdamped_Langevin} are obtained by a mere scaling of $\vec{R}^{(i)}(t)$ with the constant prefactors $\sqrt{2 k_B T / \gamma^{(i)}}$.
Therefore, the noise in BD is additive, since $g(X(t), t) = \mathrm{const.}$ in the sense of eq.\ \eqref{eq:SDE_general}.
This is a crucial property for the construction of a simple higher-order integrator below in Sec.\ \ref{sec:Application_to_BD}.

In computer simulations, particle trajectories are obtained from eq.\ \eqref{eq:overdamped_Langevin} by numerical integration.
Contrary to the numerics of ODEs, where higher-order schemes and adaptivity are textbook material, the derivation of corresponding methods for SDEs poses several difficulties which we address below.
Due to the complications, SDEs of type \eqref{eq:SDE_general} are often integrated via the Euler-Maruyama method instead, which follows the notion of the explicit Euler scheme for ODEs.
Thereby, the true solution of eq.\ \eqref{eq:SDE_general} with initial value $X(0) = X_0$ is approximated in $t \in [0, T]$ by partitioning the time interval into $n$ equidistant subintervals of length $\Delta t = T / n$.
Then, for $0 \leq k < n$, a timestep is defined by
\begin{equation}
  \label{eq:Euler-Maruyama}
  X_{k + 1} = X_k + f(X_k, t_k) \Delta t + g(X_k, t_k) \Delta W_k
\end{equation}
with Wiener increments $\Delta W_k$.
An Euler-Maruyama step is also incorporated in the adaptive BD method which we construct below, applying eq.\ \eqref{eq:Euler-Maruyama} to the overdamped Langevin equation \eqref{eq:overdamped_Langevin}.

Crucially, the random increments $\Delta W_k$ in each Euler-Maruyama step \eqref{eq:Euler-Maruyama} have to be constructed from independent and identically distributed normal random variables with expectation value $\E(\Delta W_k) = 0$ and variance $\Var(\Delta W_k) = \Delta t$.
In practice, this is realized by drawing a new random number $R$ (or vector thereof) from a pseudo-random number generator obeying the normal distribution $\mathcal{N}(0, \Delta t) = \sqrt{\Delta t} \mathcal{N}(0, 1)$ in each step $k$ and setting $\Delta W_k = R$ in eq.\ \eqref{eq:Euler-Maruyama}.
The process of obtaining such a scalar (or vectorial) random increment $R$ will be denoted in the following by
\begin{equation}
  \label{eq:normal_distribution}
  R \sim \mathcal{N}(\mu, \eta)
\end{equation}
where $\mathcal{N}(\mu, \eta)$ is a scalar (or multivariate) normal distribution with expectation value $\mu$ and variance $\eta$.

As in the case of ODEs, an important measure for the quality of an integration method is its order of convergence.
However, what convergence exactly means in the case of SDEs must be carefully reconsidered due to their stochasticity.
We refer the reader to the pertinent literature (see e.g.\ Ref.\ \onlinecite{KloedenNumericalSolutionStochastic1992}) and only summarize the key concepts and main results in the following.

Since both the approximation $X_k$ and the true solution $X(t_k)$ are random variables, one can define two distinct convergence criteria.
For a certain method with discretisation $\Delta t \rightarrow 0$, \emph{weak convergence} captures the error of average values whereas \emph{strong convergence} assesses the error of following a specific realization of a random process.
One can show that the Euler-Maruyama method has a strong convergence order of 0.5, i.e.\ when increasing or decreasing the stepsize $\Delta t$, the error of the numerical solution only scales with $\Delta t^{0.5}$.
For general $g(X(t), t)$ in eq.\ \eqref{eq:SDE_general}, the construction of schemes with higher strong order is complicated due to the occurence of higher-order stochastic integrals.
Practically, this means that the careful evaluation of additional random variables is necessary in each iteration step.
These random variables then enable the approximation of the stochastic integrals.
There exist schemes of Runge-Kutta type with strong orders up to 2, although only strong order 1 and 1.5 Runge-Kutta methods are mostly used due to practical concerns \cite{BurrageHighStrongOrder1996,RosslerRungeKuttaMethods2010}.

In order to incorporate adaptivity, one needs a means of comparison of two integration schemes of different strong order to formulate a local error estimate for the proposed step.
If the error that occurs in a specific timestep is too large (we define below precisely what we mean by that), the step is rejected and a retrial with a smaller value of $\Delta t$ is performed.
Otherwise, the step is accepted and, based on the previously calculated error, a new optimized stepsize is chosen for the next step, which hence can be larger than the previous one.
This protocol makes an optimal and automatic control of the stepsize possible, meaning that $\Delta t$ can both be reduced when the propagation of the SDE is difficult and it can be increased when the error estimate allows to do so.
Similar to the case of ODEs, it is computationally advantageous to construct so-called embedded integration schemes, analogous to e.g.\ Runge-Kutta-Fehlberg integrators \cite{FehlbergKlassischeRungeKuttaFormelnVierter1970}, which minimize the number of costly evaluations of the right-hand side of eq.\ \eqref{eq:SDE_general}.
Developments have been made in this direction e.g.\ with embedded stochastic Runge-Kutta methods \cite{RosslerEmbeddedStochasticRungeKutta2003}.

There is still one caveat to consider when rejecting a step, in that one has to be careful to preserve the properties of the Wiener process.
In the naive approach of simply redrawing new uncorrelated random increments, rejections would alter the random process implicitly.
The reason lies in the introduction of an undesired bias.
Since large random increments (generally causing larger errors) get rejected more often, the variance of the Wiener process would be systematically decreased, ultimately violating its desired properties.
To avoid this effect, it must be guaranteed that once a random increment is chosen, it will not be discarded until the time interval it originally applied to has passed.
Consequently, when rejecting a trial step and retrying the numerical propagation of the SDE with smaller time intervals, new random increments cannot be drawn independently for those substeps anymore.
The new random increments must instead be created based on the rejected original timestep such that an unbiased Brownian path is still followed.

The formal framework to the above procedure is given by the so-called Brownian bridge theorem \cite{IbeMarkovProcessesStochastic2013}, which interpolates a Wiener process between known values at two timepoints.
If $W(0) = 0$ and $W(\Delta t) = R$ are given (e.g.\ due to the previous rejection of a timestep of length $\Delta t$ where the random increment $R$ has been drawn), then a new intermediate random value must be constructed by
\begin{equation}
  \label{eq:Brownian_bridge}
  W(q \Delta t) \sim \mathcal{N}(q R, (1 - q) q \Delta t), \quad 0 < q < 1,
\end{equation}
such that the statistical properties of the to-be-simulated Wiener process are left intact and a substep $q \Delta t$ can be tried.
The value of $q$ thereby sets the fraction of the original time interval to which the Wiener process shall be interpolated.
Equation \eqref{eq:Brownian_bridge} extends naturally (i.e.\ component-wise) to the multivariate case and it hence enables the construction of interpolating random vectors in a straightforward manner.

With this idea in mind, several adaptive timestepping algorithms for SDEs have been designed \cite{GainesVariableStepSize1997, MauthnerStepSizeControl1998, BurrageVariableStepsizeImplementation2002, LambaAdaptiveTimesteppingAlgorithm2003, BurrageAdaptiveStepsizeBased2004, LambaAdaptiveEulerMaruyamaScheme2006, SotiropoulosAdaptiveTimeStep2008}.
Still, most of these approaches are quite restrictive in the choice of timesteps (e.g.\ only allowing discrete variations such as halving or doubling) \cite{GainesVariableStepSize1997, SotiropoulosAdaptiveTimeStep2008}, involve the deterministic or random part only separately into an a-priori error estimation \cite{LambaAdaptiveTimesteppingAlgorithm2003, LambaAdaptiveEulerMaruyamaScheme2006, IlieVariableTimesteppingPathwise2012}, or store rejected timesteps in a costly manner, e.g.\ in the form of Brownian trees \cite{GainesVariableStepSize1997}.
In particular, the above methods rely on precomputed Brownian paths and do not illustrate an ad hoc generation, which is desirable from a performance and memory consumption standpoint in a high-dimensional setting like BD.

In contrast, a very flexible and performant class of adaptive timestepping algorithms called Rejection Sampling with Memory (RSwM) has recently been developed by \citeauthor{RackauckasAdaptiveMethodsStochastic2017} \cite{RackauckasAdaptiveMethodsStochastic2017}.
Their work provides the arguably optimal means for the adaptive numerical solution of SDEs while still being computationally efficient in the generation of random increments as well as in the handling of rejections.
We therefore use RSwM in the construction of an adaptive algorithm and specialize the method to Brownian motion in the following.

\section{Application to Brownian Dynamics}
\label{sec:Application_to_BD}
Based on the remarks of Sec.\ \ref{sec:Numerics_SDEs}, we next proceed to apply the general framework to the case of BD with the overdamped Langevin equation \eqref{eq:overdamped_Langevin} forming the underlying SDE.
An embedded integration scheme is constructed which allows the derivation of an error estimate and an acceptance criterion in each step.
Furthermore, the application of RSwM for handling rejected timesteps in BD is shown and discussed.
We also illustrate how the calculation of observables from sampling of phase space functions has to be altered in a variable timestep scenario.

\subsection{Embedded Heun-Euler method}
\label{subsec:Embedded_Heun-Euler}
Regarding the overdamped Langevin equation \eqref{eq:overdamped_Langevin}, a major simplification exists compared to the general remarks made in Sec.\ \ref{sec:Numerics_SDEs}.
Due to the noise term being trivial, some higher-order schemes can be constructed by only evaluating the deterministic forces for different particle configurations.
Crucially, no iterated stochastic integrals are needed, which would have to be approximated in general higher strong-order integrators by using additional random variables \cite{RosslerRungeKuttaMethods2010}.
In the following, we apply a scheme similar to the one suggested by \citeauthor{LambaAdaptiveEulerMaruyamaScheme2006} \cite{LambaAdaptiveEulerMaruyamaScheme2006} for general SDEs to eq.\ \eqref{eq:overdamped_Langevin} and term it \emph{embedded Heun-Euler method} due to its resemblance to the corresponding ODE integrators.
Two different approximations $\bar{\vec{r}}^N_{k + 1}$ and $\vec{r}^N_{k + 1}$ are calculated in each trial step by
\begin{align}
  \label{eq:embedded_Heun-Euler_Euler}
  \bar{\vec{r}}^{(i)}_{k + 1} &= \vec{r}^{(i)}_k + \frac{1}{\gamma^{(i)}} \vec{F}^{(i)}(\vec{r}^N_k) \Delta t_k + \sqrt{\frac{2 k_B T}{\gamma^{(i)}}} \vec{R}^{(i)}_k,\\
  \label{eq:embedded_Heun-Euler_Heun}
  \begin{split}
    \vec{r}^{(i)}_{k + 1} &= \vec{r}^{(i)}_k + \frac{1}{2 \gamma^{(i)}} \left( \vec{F}^{(i)}(\vec{r}^N_k) + \vec{F}^{(i)}(\bar{\vec{r}}^N_{k + 1}) \right) \Delta t_k\\
    &\qquad + \sqrt{\frac{2 k_B T}{\gamma^{(i)}}} \vec{R}^{(i)}_k.
  \end{split}
\end{align}

Equation \eqref{eq:embedded_Heun-Euler_Euler} is the conventional Euler-Maruyama step, and hence constitutes the application of eq.\ \eqref{eq:Euler-Maruyama} to the overdamped Langevin equation \eqref{eq:overdamped_Langevin}.
Equation \eqref{eq:embedded_Heun-Euler_Heun} resembles the second order Heun algorithm or midpoint scheme for ODEs \cite{StuartDynamicalSystemsNumerical1996} and has been formally derived for SDEs in the context of stochastic Runge-Kutta methods \cite{HoneycuttStochasticRungeKuttaAlgorithms1992}.
Since the deterministic forces $\vec{F}^N$ are evaluated at the initial particle configuration $\vec{r}_k^N$ both in eq.\ \eqref{eq:embedded_Heun-Euler_Euler} and eq.\ \eqref{eq:embedded_Heun-Euler_Heun}, we have constructed an embedded integration method.
This is favorable regarding computational cost, since the numerical result of $\vec{F}^N(\vec{r}_k^N)$ is evaluated once in eq.\ \eqref{eq:embedded_Heun-Euler_Euler} and reused in eq.\ \eqref{eq:embedded_Heun-Euler_Heun}.
Only one additional computation of the deterministic forces at the intermediate particle configuration $\bar{\vec{r}}_{k+1}^N$ is then needed in the Heun step \eqref{eq:embedded_Heun-Euler_Heun}.

In each trial step, the same realization of random vectors $\vec{R}^N_k$ must be used in both eqs.\ \eqref{eq:embedded_Heun-Euler_Euler} and \eqref{eq:embedded_Heun-Euler_Heun}.
Recall that the random displacements have to obey the properties of multivariate Wiener increments to model the non-deterministic forces.
In conventional BD with fixed stepsize $\Delta t_k = \Delta t = \mathrm{const.}$, these random vectors can therefore be drawn independently via $\vec{R}^{(i)}_k \sim \mathcal{N}(0, \Delta t)$ for each particle $i$.
If rejections of trial steps are possible, $\vec{R}^{(i)}_k$ must instead be constructed as described in Sec.\ \ref{subsec:RSwM_BD}.

When the embedded Heun-Euler step \eqref{eq:embedded_Heun-Euler_Euler} and \eqref{eq:embedded_Heun-Euler_Heun} is applied to BD, the improvement over the conventional method is twofold.
Firstly, the Heun step \eqref{eq:embedded_Heun-Euler_Heun} can be used as a better propagation method as already analyzed by \citeauthor{FixmanImplicitAlgorithmBrownian1986} \cite{FixmanImplicitAlgorithmBrownian1986} and \citeauthor{IniestaSecondOrderAlgorithm1990} \cite{IniestaSecondOrderAlgorithm1990}.
Several further higher-order schemes, mostly of Runge-Kutta type, have been used in BD simulations \cite{HoneycuttStochasticRungeKuttaAlgorithms1992,HeyesMoreEfficientBrownian2000}.
These methods often lead to increased accuracy and even efficiency due to bigger timesteps becoming achievable, which outweighs the increased computational cost per step.
Since the prefactor of the random force is trivial (i.e.\ constant) in the overdamped Langevin equation \eqref{eq:overdamped_Langevin}, higher strong-order schemes are easily constructable.
This situation stands in contrast to the more complicated SDEs of type \eqref{eq:SDE_general} with general noise term $g(X(t), t)$.

Secondly, with two approximations of different order at hand, assessing their discrepancy allows to obtain an estimate of the discretisation error in each step, which is a fundamental prerequisite in the construction of adaptive timestepping algorithms.
For this, we exploit the additive structure of the noise term in eq.\ \eqref{eq:overdamped_Langevin} again and recognize that such an error estimate can be obtained by only comparing the deterministic parts of eqs.\ \eqref{eq:embedded_Heun-Euler_Euler} and \eqref{eq:embedded_Heun-Euler_Heun}.
Nevertheless, note that the random displacements are already contained in $\bar{\vec{r}}^N_{k + 1}$.
This makes the deterministic part of eq.\ \eqref{eq:embedded_Heun-Euler_Heun} implicitly dependent on the realization of $\vec{R}^N_k$, which is opposed to Ref.\ \onlinecite{LambaAdaptiveEulerMaruyamaScheme2006} where an error estimate is defined without involving the random increments at all.

At a given step $k \rightarrow k + 1$, one can construct the automatic choice of an appropriate $\Delta t_k$.
For this, the error of a trial step with length $\Delta t$ is evaluated.

We define a particle-wise error
\begin{equation}
  \label{eq:E}
  E^{(i)} = \lVert \Delta \bar{\vec{r}}^{(i)} - \Delta \vec{r}^{(i)} \rVert = \frac{\Delta t}{2 \gamma^{(i)}} \lVert \vec{F}^{(i)}(\bar{\vec{r}}^N_{k + 1}) - \vec{F}^{(i)}(\vec{r}^N_k) \rVert
\end{equation}
with $\Delta \bar{\vec{r}}^{(i)} = \bar{\vec{r}}^{(i)}_{k + 1} - \vec{r}^{(i)}_k$ and $\Delta \vec{r}^{(i)} = \vec{r}^{(i)}_{k + 1} - \vec{r}^{(i)}_k$.
For each particle $i$, this error is compared to a tolerance
\begin{equation}
  \label{eq:tol}
  \tau^{(i)} = \epsilon_\mathrm{abs} + \epsilon_\mathrm{rel} \lVert \Delta \vec{r}^{(i)} \rVert
\end{equation}
consisting of an absolute and a relative part with user-defined coefficients $\epsilon_\mathrm{abs}$ and $\epsilon_\mathrm{rel}$.
Note that $\Delta \vec{r}^{(i)}$ is used in eq.\ \eqref{eq:tol}, since it captures the true particle displacement more accurately than $\Delta \bar{\vec{r}}^{(i)}$ due to its higher order.
Additionally, we stress that $\Delta \vec{r}^{(i)}$ decreases on average for shorter timesteps, such that $\tau^{(i)}$ indirectly depends on the trial $\Delta t$, which limits the accumulation of errors after multiple small steps.

Then a total error estimate
\begin{equation}
  \label{eq:e}
  \mathcal{E} = \left\lVert \left( \frac{E^{(i)}}{\tau^{(i)}} \right)_{1 \leq i \leq N} \right\rVert
\end{equation}
can be calculated for the trial step.
While the Euclidean norm is the canonical choice in eq.\ \eqref{eq:E}, there is a certain freedom of choosing an appropriate norm $\lVert \cdot \rVert$ in eq.\ \eqref{eq:e}.
The $2$- or $\infty$-norm defined respectively as
\begin{align}
  \left\lVert (x^{(i)})_{1 \leq i \leq N} \right\rVert_{2} &= \sqrt{\frac{1}{N} \sum_{i = 1}^N | x^{(i)} |^2},\\
  \left\lVert (x^{(i)})_{1 \leq i \leq N} \right\rVert_{\infty} &= \max_{1 \leq i \leq N} | x^{(i)} |,
\end{align}
may both come up as natural and valid options (note that we normalise the standard 2-norm by $\sqrt{N}$ to obtain an intensive quantity).
However, in eq.\ \eqref{eq:e}, where a reduction from particle-wise errors to a global scalar error takes place, this has crucial implications to the kind of error that is being controlled.
If the $2$-norm is used, then $\epsilon_\mathrm{abs}$ and $\epsilon_\mathrm{rel}$ set the \emph{mean} absolute and relative tolerance for all particles.
In practice for large particle number $N$, this can lead to substantial single-particle errors becoming lost in the global average.
Therefore, it is advisable to use the $\infty$-norm for the reduction in eq.\ \eqref{eq:e} to be able to set a \emph{maximum single-particle} absolute and relative tolerance, i.e.\ if $\mathcal{E} < 1$, $E^{(i)} < \tau^{(i)}$ for all $i = 1, \dots, N$.

Following the design of adaptive ODE solvers and ignoring stochasticity, an expansion of an embedded pair of methods with orders $p$ and $p - 1$ shows that an error estimate of type \eqref{eq:e} is of order $p$.
Thus, a timestep of length $q \Delta t$ with
\begin{equation}
  \label{eq:q_ODE}
  q = \mathcal{E}^{- \frac{1}{p}}
\end{equation}
could have been chosen to marginally satisfy the tolerance requirement $\mathcal{E} < 1$.

Considering the recommendation of Ref.\ \onlinecite{RackauckasAdaptiveMethodsStochastic2017} which discusses the application of such a timestep scaling factor to embedded methods for SDEs, we set
\begin{equation}
  \label{eq:q}
  q = \left( \frac{1}{\alpha \mathcal{E}} \right)^2.
\end{equation}
Here both a more conservative exponent is chosen than in eq.\ \eqref{eq:q_ODE} and also a safety factor $\alpha = 2$ is introduced, as we want to account for stochasticity and for the low order of our integrators, which results in a low order of the error estimate \eqref{eq:e}.

With the choice \eqref{eq:q}, one can distinguish two possible scenarios in each trial step:
\begin{itemize}
  \item $q \geq 1$: Accept the trial step, i.e.\ set $\Delta t_k = \Delta t$ and advance the particle positions with eq.\ \eqref{eq:embedded_Heun-Euler_Heun}, and then continue with $k + 1 \rightarrow k + 2$.
  \item $q < 1$: Reject the trial step and retry $k \rightarrow k + 1$ with a smaller timestep.
\end{itemize}
In both cases, the timestep is adapted afterwards via $\Delta t \gets q \Delta t$.
Here and in the following, the notation $a \gets b$ denotes an assignment of the value $b$ to the variable $a$.

It is advisable to restrict the permissible range of values for $q$ by defining lower and upper bounds $q_\mathrm{min} \leq q \leq q_\mathrm{max}$ such that the adaptation of $\Delta t$ is done with
\begin{equation}
  \label{eq:q_bounded}
  q \gets \min(q_\mathrm{max}, \max(q_\mathrm{min}, q))
\end{equation}
While commonly chosen as $q_\mathrm{max} \approx 10$ and $q_\mathrm{min} \approx 0.2$ for ODE solvers, due to stochasticity and the possibility of drawing ``difficult'' random increments, $q_\mathrm{min}$ should certainly be decreased in the case of SDEs to avoid multiple re-rejections.
Vice versa, a conservative choice of $q_\mathrm{max}$ prevents an overcorrection of the timestep in case of ``fortunate'' random events.
We set $q_\mathrm{max} = 1.2$ and $q_\mathrm{min} = 0.001$ in practical applications to achieve a rapid adaptation in case of rejected trial steps and a careful approach to larger timesteps after accepted moves.

One can also impose limits for the range of values of $\Delta t$, such as a maximum bound $\Delta t_\mathrm{max} \geq \Delta t$.
Restriction by a minimum value $\Delta t_\mathrm{min} \leq \Delta t$, however, could lead to a forced continuation of the simulation with an actual local error that lies above the user-defined tolerance.
Thus, this is not recommended.
In our test cases described below, we see no need to restrict the timestep as the adaptive algorithm does not show unstable behavior without such a restriction.

Most concepts of this section can be generalized in a straightforward manner to non-overdamped Langevin dynamics, where particle inertia is explicitly considered in the stochastic equations of motion.
However, the embedded Heun-Euler method \eqref{eq:embedded_Heun-Euler_Euler} and \eqref{eq:embedded_Heun-Euler_Heun} might not be a suitable integration scheme in this case.
We therefore outline the modifications that are necessary in the construction of an adaptive algorithm for general Langevin dynamics in Appendix \ref{appendix:general_Langevin}.

\subsection{Rejection Sampling with Memory in BD}
\label{subsec:RSwM_BD}

\begin{figure}[htbp]
  \centering
  \input{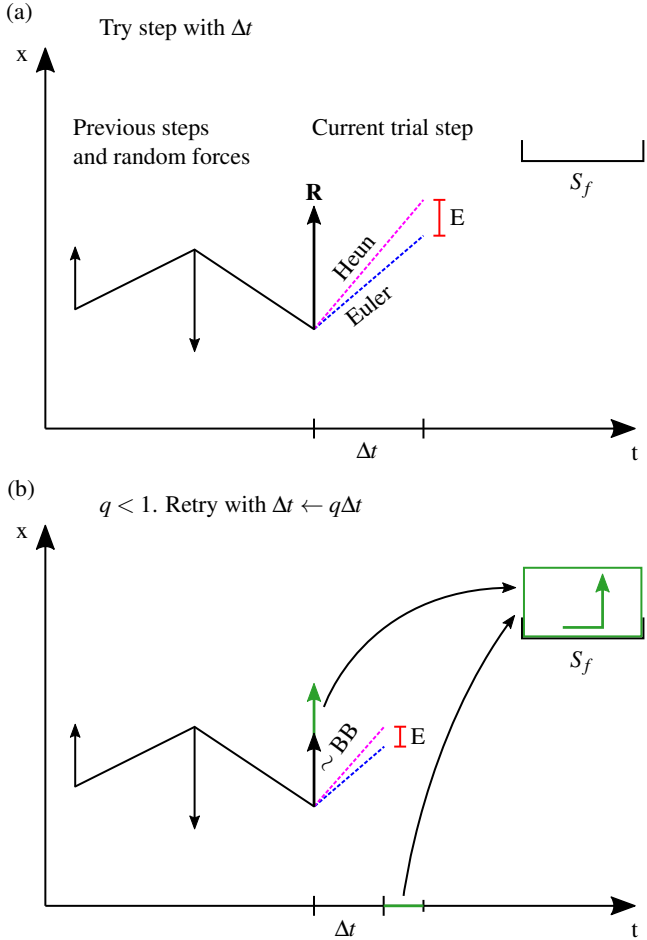}
  \caption{The trial step is rejected (a) and a retrial with a smaller value of $\Delta t$ is performed (b) if the discrepancy between eq.\ \eqref{eq:embedded_Heun-Euler_Euler} and eq.\ \eqref{eq:embedded_Heun-Euler_Heun} is large. To preserve the properties of the Brownian motion, the Brownian bridge (BB) theorem \eqref{eq:Brownian_bridge} is used to interpolate the random process at the intermediate timepoint $q \Delta t$. The difference between the bridged random sample and the rejected original random sample is stored along with the remaining time difference onto the stack $S_f$. This is indicated in (b), where $S_f$ now contains one element that holds the residual time interval (horizontal segment) and random increment (vertical arrow). Thus, in future steps, the Brownian path can be reconstructed from elements on $S_f$ and the properties of the Wiener process remain intact. Note that for correctness in case of re-rejections, one requires a second stack $S_u$ as explained below and in fig.\ \ref{fig:stacks}.}
  \label{fig:rejection}
\end{figure}

\begin{figure}[htbp]
  \centering
\begingroup%
  \makeatletter%
  \providecommand\color[2][]{%
    \errmessage{(Inkscape) Color is used for the text in Inkscape, but the package 'color.sty' is not loaded}%
    \renewcommand\color[2][]{}%
  }%
  \providecommand\transparent[1]{%
    \errmessage{(Inkscape) Transparency is used (non-zero) for the text in Inkscape, but the package 'transparent.sty' is not loaded}%
    \renewcommand\transparent[1]{}%
  }%
  \providecommand\rotatebox[2]{#2}%
  \newcommand*\fsize{\dimexpr\f@size pt\relax}%
  \newcommand*\lineheight[1]{\fontsize{\fsize}{#1\fsize}\selectfont}%
  \ifx\svgwidth\undefined%
    \setlength{\unitlength}{242.63999176bp}%
    \ifx\svgscale\undefined%
      \relax%
    \else%
      \setlength{\unitlength}{\unitlength * \real{\svgscale}}%
    \fi%
  \else%
    \setlength{\unitlength}{\svgwidth}%
  \fi%
  \global\let\svgwidth\undefined%
  \global\let\svgscale\undefined%
  \makeatother%
  \begin{picture}(1,0.74183979)%
    \lineheight{1}%
    \setlength\tabcolsep{0pt}%
    \put(0,0){\includegraphics[width=\unitlength,page=1]{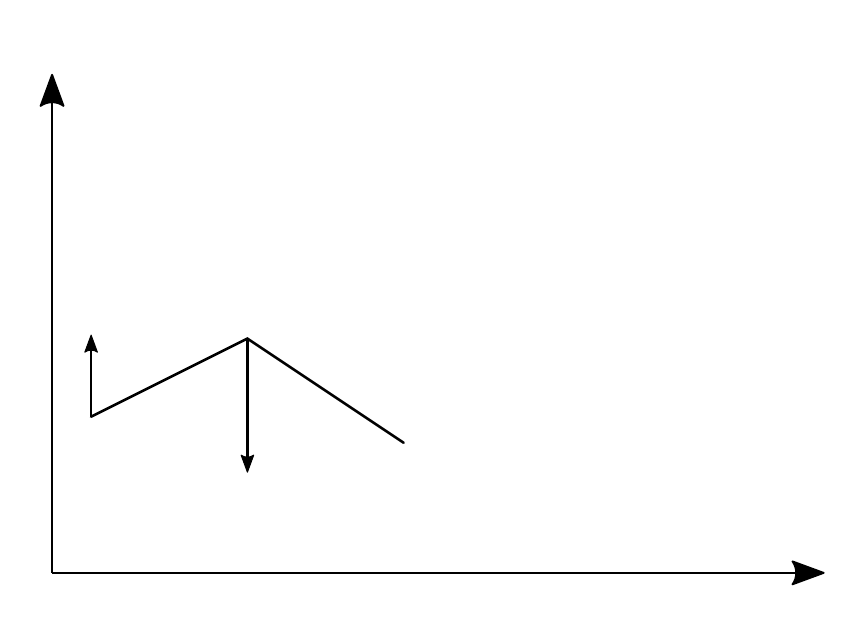}}%
    \put(0.97233659,0.01354731){\makebox(0,0)[lt]{\lineheight{1.25}\smash{\begin{tabular}[t]{l}t\end{tabular}}}}%
    \put(0.01637265,0.63365482){\makebox(0,0)[lt]{\lineheight{1.25}\smash{\begin{tabular}[t]{l}x\end{tabular}}}}%
    \put(0,0){\includegraphics[width=\unitlength,page=2]{setup_next_step_with_gap.pdf}}%
    \put(0.89525473,0.43338582){\makebox(0,0)[t]{\lineheight{1.25}\smash{\begin{tabular}[t]{c}$S_f$\end{tabular}}}}%
    \put(0,0){\includegraphics[width=\unitlength,page=3]{setup_next_step_with_gap.pdf}}%
    \put(0.60892682,0.44510389){\rotatebox{44.856525}{\makebox(0,0)[lt]{\lineheight{1.25}\smash{\begin{tabular}[t]{l}$\sim \mathcal{N}(0, \Delta t_\mathrm{gap})$\end{tabular}}}}}%
    \put(0.68620179,0.09272997){\makebox(0,0)[t]{\lineheight{1.25}\smash{\begin{tabular}[t]{c}$\Delta t_\mathrm{gap}$\end{tabular}}}}%
    \put(0.65838282,0.01545501){\makebox(0,0)[t]{\lineheight{1.25}\smash{\begin{tabular}[t]{c}$\Delta t$\end{tabular}}}}%
  \end{picture}%
\endgroup%

  \caption{After an accepted step has been performed, a new random increment is prepared for the next trial step of length $\Delta t$. If available, future information -- stemming from previously rejected trial steps -- has to be incorporated in the generation of new random vectors in order to retain the properties of Brownian motion. In the shown case, the future information stack contains one element, which is popped and accumulated to the new random increment and time interval. The stack is now empty and a difference $\Delta t_\mathrm{gap}$ to the goal timestep $\Delta t$ remains. For this gap, a new uncorrelated random increment has to be generated from $\mathcal{N}(0, \Delta t_\mathrm{gap})$ to complete the preparation of the next trial step.}
  \label{fig:setup_next_step_with_gap}
\end{figure}

\begin{figure}[htbp]
  \centering
  \input{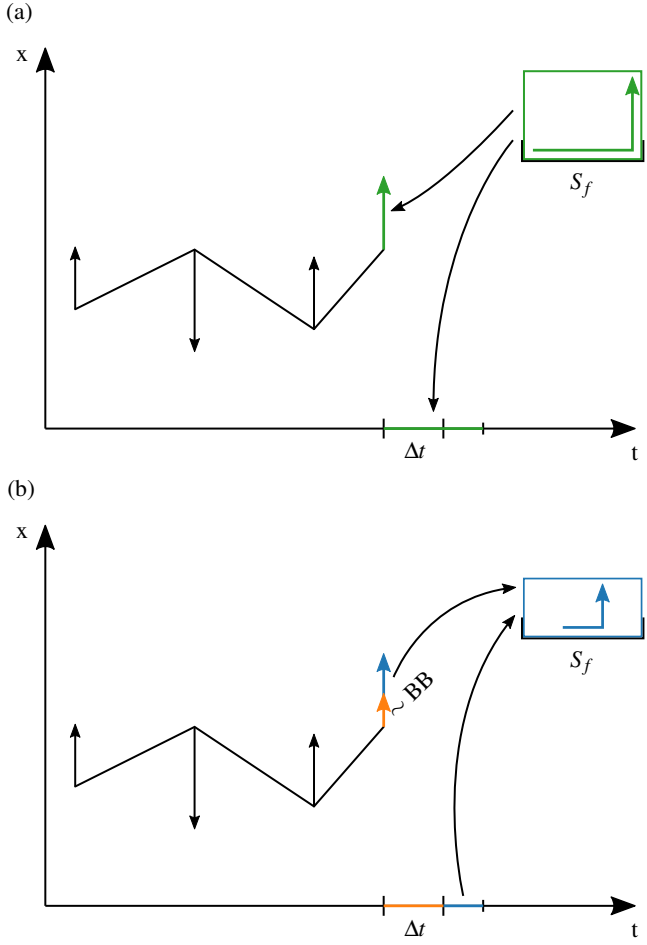}
  \caption{The situation is similar to fig.\ \ref{fig:setup_next_step_with_gap}, where a step is rejected (a) and then retried (b). Here, the future information reaches further than the goal timestep $\Delta t$. In this case, the Brownian bridge (BB) has to be applied for the interpolation of an element from the future information stack. Generally, one pops the elements of the future stack $S_f$ one after another and accumulates the random increments and time intervals until the element which crosses $\Delta t$ is reached, to which the bridging theorem is then applied. In the shown example, the interpolation is done with the first element of $S_f$, since it already surpasses $\Delta t$. Similar to fig.\ \ref{fig:rejection}, the remainder of the bridged increment and time interval is pushed back onto $S_f$.}
  \label{fig:setup_next_step_with_bridge}
\end{figure}

\begin{figure}[htbp]
  \centering
  \input{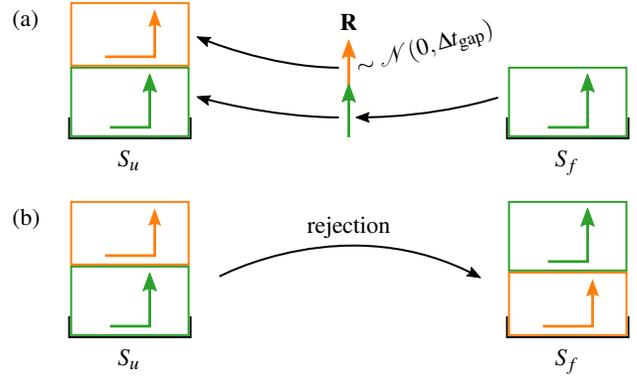}
  \caption{To keep track of the random increments that are used in the construction of $\vec{R}$, a second stack $S_u$ is introduced. In (a), the situation of fig.\ \ref{fig:setup_next_step_with_gap} is shown as an example, where $\vec{R}$ consists of an element of $S_f$ (stemming from a previous rejection) and a Gaussian contribution $\vec{R}_\mathrm{gap} \sim \mathcal{N}(0, \Delta t_\mathrm{gap})$. In case of rejection, the elements that were popped from $S_f$ as well as newly drawn increments such as $\vec{R}_\mathrm{gap}$ would be lost unrecoverably. By using $S_u$ as an intermediate storage, all contributions to $\vec{R}$ can be transferred back to $S_f$ such that no information about drawn random increments is lost and the same Brownian path is still followed. These considerations apply as well when random vectors are drawn via the Brownian bridge theorem, as e.g.\ in figs.\ \ref{fig:rejection} and \ref{fig:setup_next_step_with_bridge}.}
  \label{fig:stacks}
\end{figure}

We still have to prescribe the generation of the random vectors $\vec{R}^{N}_k$ that appear both in eqs.\ \eqref{eq:embedded_Heun-Euler_Euler} and \eqref{eq:embedded_Heun-Euler_Heun}.
For clarity, we consider a single trial step and denote the set of corresponding random increments with $\vec{R}$ (the sub- and superscript is dropped).
Obviously, $\vec{R}$ can no longer be chosen independently in general but has to incorporate previously rejected timesteps as long as they are relevant, i.e.\ as long as their original time interval has not passed yet.
For this, we apply the Rejection Sampling with Memory (RSwM) algorithm to BD.
\citeauthor{RackauckasAdaptiveMethodsStochastic2017} \cite{RackauckasAdaptiveMethodsStochastic2017} describe three variants of RSwM which they refer to as RSwM1, RSwM2, and RSwM3, and which differ in generality of the possible timesteps and algorithmic complexity.
We aim to reproduce RSwM3 because it enables optimal timestepping and still ensures correctness in special but rare cases such as re-rejections.

Common to all of the RSwM algorithms is storing parts of rejected random increments onto a stack $S_f$.
We refer to elements of this stack as \emph{future information}, since they have to be considered in the construction of future steps.
This becomes relevant as soon as a step is rejected due to a too large error.
Then a retrial is performed by decreasing $\Delta t$ and drawing bridging random vectors via eq.\ \eqref{eq:Brownian_bridge}.
The difference between rejected and bridging random vectors must not be forgotten but rather be stored on the future information stack, cf.\ fig.\ \ref{fig:rejection}.

On the other hand, if a step is accepted, new random vectors have to be prepared for the next time interval $\Delta t$.
If no future information is on the stack, this can be done conventionally via drawing Gaussian distributed random vectors according to $\vec{R} \sim \mathcal{N}(0, \Delta t)$.
If the stack $S_f$ is not empty and thus future information is available, then elements of the stack are popped one after another, i.e.\ they are taken successively from the top.
The random vectors as well as the time interval of each popped element are accumulated in temporary variables which then hold the sum of those respective time intervals and random vectors.
The stack could be empty before the accumulated time reaches $\Delta t$, i.e.\ there could still be a difference $\Delta t_\mathrm{gap}$.
In this case, one draws new random vectors $\vec{R}_\mathrm{gap} \sim \mathcal{N}(0, \Delta t_\mathrm{gap})$ to compensate for the difference and adds them to the accumulated ones, cf.\ fig. \ref{fig:setup_next_step_with_gap}, before attempting the trial step.
Otherwise, if the future information reaches further than $\Delta t$, then there is one element that passes $\Delta t$.
One takes this element, splits it in ``before $\Delta t$'' and ``after $\Delta t$'' and draws bridging random vectors for ``before $\Delta t$'' according to eq.\ \eqref{eq:Brownian_bridge} which are again added to the accumulated ones.
The rest of this element (``after $\Delta t$'') can be pushed back to the future information stack and the step $\Delta t$ can be tried with the accumulated vectors set as $\vec{R}$ in eqs.\ \eqref{eq:embedded_Heun-Euler_Euler} and \eqref{eq:embedded_Heun-Euler_Heun}, cf.\ fig.\ \ref{fig:setup_next_step_with_bridge}.

At this stage, we have constructed RSwM2 for BD, which is not capable of handling all edge cases yet as pointed out in Ref.\ \onlinecite{RackauckasAdaptiveMethodsStochastic2017}.
If future information is popped from the stack to prepare $\vec{R}$ for the next step, and this next step is then rejected, we have lost all popped information unrecoverably.
To circumvent this, one adds a second stack $S_u$ that stores information which is currently used for the construction of $\vec{R}$.
We refer to elements of this stack as \emph{information in use}.
If a step is rejected, the information in use can be moved back to the future information stack so that no elements are lost in multiple retries, cf.\ fig.\ \ref{fig:stacks}.
With this additional bookkeeping, correctness and generality is ensured in all cases and the RSwM3 algorithm is complete.

Notably, with the structuring of information into stacks where only the top element is directly accessible (``last in first out''), the chronological time order is automatically kept intact so that one only has to store time intervals and no absolute timepoints.
Furthermore, searching or sorting of elements is prevented entirely which makes all operations $\mathcal{O}(1)$ and leads to efficient implementations.

We point out that the original RSwM3 rejection branch as given in Ref.\ \onlinecite{RackauckasAdaptiveMethodsStochastic2017} was not entirely correct and draw a comparison to our rectifications in Appendix \ref{appendix:RSwM3_correctness}, which have been brought to attention \footnote{This issue has been discussed in \url{https://github.com/SciML/DiffEqNoiseProcess.jl/pull/80}.} and have since been fixed in the reference implementation DifferentialEquations.jl \cite{DifferentialEquations.jl-2017}.
Crucially, the correction not only applies to the case of BD, but it rather is relevant for the solution of general SDEs as well.
A full pseudocode listing of one adaptive BD trial step utilizing RSwM3 is given in alg.\ \ref{alg:RSwM3_BD} in Appendix \ref{appendix:RSwM3_BD} along with further explanation of technical details.\footnote{The corresponding implementation in C++ can be found in \url{https://gitlab.uni-bayreuth.de/bt306964/mbd}} 

\subsection{Sampling of observables}
\label{subsec:Sampling_observables}
Within BD, observables can be obtained from the sampling of configuration space functions.
As an example, consider the one-body density profile $\rho(\vec{r})$, which is defined as the average of the density operator $\hat{\rho}(\vec{r}, \vec{r}^N) = \sum_{i = 1}^N \delta(\vec{r} - \vec{r}_i)$.

In simulations of equilibrium or steady states, one can use a time-average over a suitably long interval $[0, T]$ to measure such quantities, i.e.\ for a general operator $\hat{A}(X, \vec{r}^N)$,
\begin{equation}
  \label{eq:Sampling_timeaverage_integral}
  A(X) \approx \frac{1}{T} \int_0^T \diff t \hat{A}(X, \vec{r}^N(t)).
\end{equation}
Note that the remaining dependence on $X$ can consist of arbitrary scalar or vectorial quantities or also be empty.
For example, $X = (\vec{r}, \vec{r}')$ or $X = \vec{r}$ for general two- or one-body fields, $X = r$ for the isotropic radial distribution function or $X = \emptyset$ for bulk quantities such as pressure or heat capacity.

Practically, $\hat{A}(X, \vec{r}^N)$ is evaluated in each step and an $X$-resolved histogram is accumulated which yields $A(X)$ after normalization.
Considering the numerical discretisation of $[0, T]$ into $n$ timesteps of constant length $\Delta t$ within a conventional BD simulation, eq.\ \eqref{eq:Sampling_timeaverage_integral} is usually implemented as
\begin{equation}
  \label{eq:Sampling_timeaverage_fixed}
  A(X) \approx \frac{1}{n} \sum_{k = 1}^n \hat{A}(X, \vec{r}_k^N).
\end{equation}

In adaptive BD with varying timestep length, one cannot use eq.\ \eqref{eq:Sampling_timeaverage_fixed} directly, since this would cause disproportionately many contributions to $A(X)$ from small timesteps and would thus lead to a biased evaluation of any observable.
Formally, the quadrature in eq.\ \eqref{eq:Sampling_timeaverage_integral} now has to be evaluated numerically at non-equidistant integration points.

Therefore, if the time interval $[0, T]$ is discretized into $n$ non-equidistant timepoints $0 = t_1 < t_2 < \dots < t_n < t_{n+1} = T$ and $\Delta t_k = t_{k + 1} - t_k$,
\begin{equation}
  \label{eq:Sampling_timeaverage_general}
  A(X) \approx \frac{1}{T} \sum_{k = 1}^n \Delta t_k \hat{A}(X, \vec{r}_k^N)
\end{equation}
constitutes a generalization of eq.\ \eqref{eq:Sampling_timeaverage_fixed} for this case that enables a straightforward sampling of observables within adaptive BD.

An alternative technique can be employed in scenarios where the state of the system shall be sampled on a sparser time grid than the one given by the integration timesteps.
Then, regular sampling points can be defined that must be hit by the timestepping procedure (e.g.\ by artificially shortening the preceding step).
On this regular time grid, eq.\ \eqref{eq:Sampling_timeaverage_fixed} can be used again.
Especially in non-equilibrium situations and for the measurement of time-dependent correlation functions such as the van Hove two-body correlation function \cite{Yeomans-ReynaSelfconsistentTheoryCollective2003,ArcherDynamicsInhomogeneousLiquids2007,HopkinsVanHoveDistribution2010,StopperStructuralRelaxationDiffusion2016,StopperBulkDynamicsBrownian2018,TreffenstadtUniversalityDrivenEquilibrium2021}, this method might be beneficial since quantities can still be evaluated at certain timepoints rather than having to construct a time-resolved histogram consisting of finite time intervals.
Note however, that timestepping to predefined halting points is not yet considered in alg.\ \ref{alg:RSwM3_BD}.

\section{Simulation results}
\label{sec:Simulation_results}
To test and illustrate the adaptive BD algorithm, we investigate the truncated and shifted Lennard-Jones fluid with interaction potential
\begin{equation}
  \Phi_\mathrm{LJTS}(r) = \begin{cases}
    \Phi_\mathrm{LJ}(r) - \Phi_\mathrm{LJ}(r_c), &r < r_c\\
    0, &r \geq r_c
  \end{cases}
\end{equation}
where
\begin{equation}
  \Phi_\mathrm{LJ}(r) = 4 \epsilon \left[ \left( \frac{\sigma}{r} \right)^{12} - \left( \frac{\sigma}{r} \right)^6 \right]
\end{equation}
and $r$ is the interparticle distance.
We set a cutoff radius of $r_c = 2.5 \sigma$ throughout the next sections and use reduced Lennard-Jones units which yield the reduced timescale $\tau = \sigma^2 \gamma / \epsilon$.

A common problem in conventional BD simulations is the choice of an appropriate timestep.
Obviously, a too small value of $\Delta t$ -- while leading to accurate trajectories -- has a strong performance impact, hindering runs which would reveal long time behavior and prohibiting extensive sampling periods, which are desirable from the viewpoint of the time average \eqref{eq:Sampling_timeaverage_general}.
Still, $\Delta t$ must be kept below a certain threshold above which results might be wrong or the simulation becomes unstable.
Unfortunately, due to the absence of any intrinsic error estimates, judgement of a chosen $\Delta t$ is generally not straightforward.
For instance, one can merely observe the stability of a single simulation run and accept $\Delta t$ if sensible output is produced and certain properties of the system (such as its energy) are well-behaved.
Another possibility is the costly conduction of several simulation runs with differing timesteps, thereby cross-validating gathered results.
Consequently, a true a priori choice of the timestep is not possible in general and test runs cannot always be avoided.

With adaptive timestepping, this problem is entirely prevented as one does not need to make a conscious choice for $\Delta t$ at all.
Instead, the maximum local error of a step is restricted by the user-defined tolerance \eqref{eq:tol}, ensuring correctness of results up to a controllable discretisation error.
This does come at the moderate cost of overhead due to the additional operations per step necessary in the embedded Heun-Euler method and the RSwM algorithm.
However, as we demonstrate in the following, the benefits of this method far outweigh the cost even in simple situations.

\subsection{Lennard-Jones bulk fluid in equilibrium}
In the following, we compare results from conventional BD to those obtained with adaptive BD.
We first consider a bulk system of size $7 \times 7 \times 6 \sigma^3$ with periodic boundary conditions at temperature $k_B T = 0.8 \epsilon$ consisting of $N = 100$ Lennard-Jones particles initialized on a simple cubic lattice.
In the process of equilibration, a gaseous and a liquid phase emerge, and the system therefore becomes inhomogeneous.

With non-adaptive Euler-Maruyama BD, a timestep $\Delta t_\mathrm{fix} = 10^{-4} \tau$ is chosen to consistently converge to this state.
This value is small enough to avoid severe problems which occur reproducibly for $\Delta t_\mathrm{fix} \gtrsim 5 \cdot 10^{-4} \tau$, where the simulation occasionally crashes or produces sudden jumps in energy due to faulty particle displacements.

\begin{figure}[htbp]
  \centering
  \includegraphics{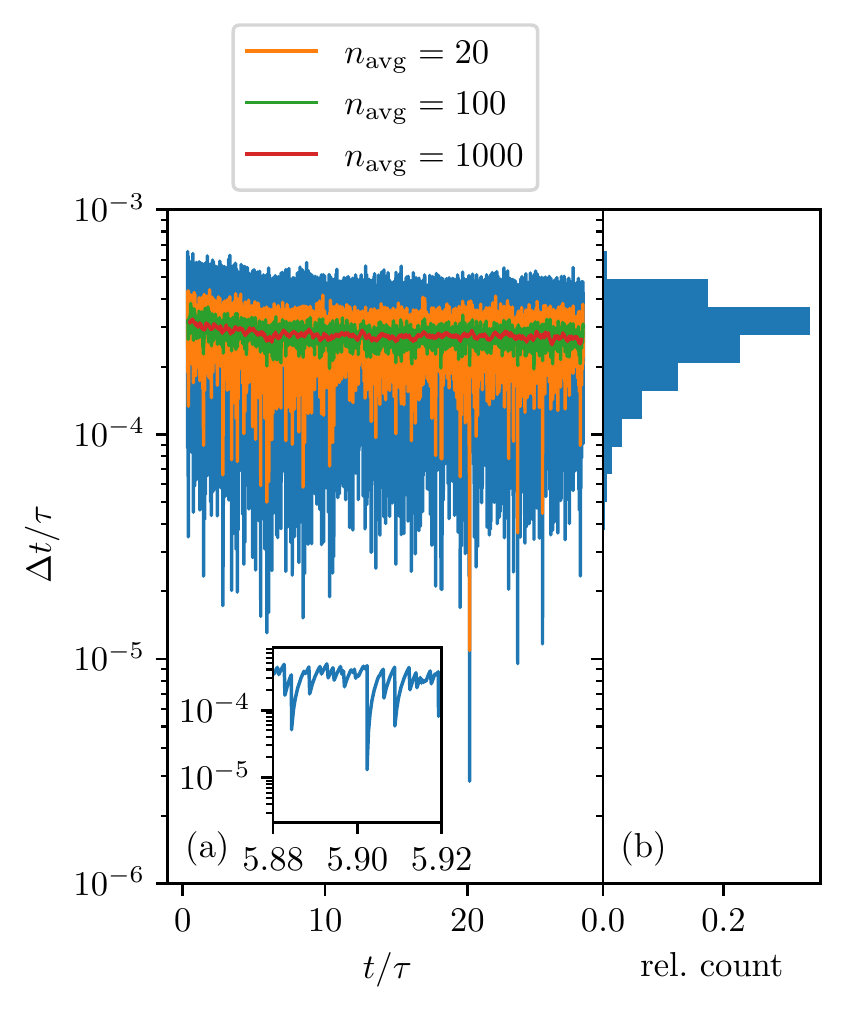}
  \caption{The evolution of chosen timesteps for accepted moves is shown in (a). To accentuate the distribution of values of $\Delta t$ further, moving averages taken over the surrounding $n_\mathrm{avg}$ points of a respective timestep record are depicted. One can see that the timestep $\Delta t$ varies rapidly in a broad range between $\Delta t \approx 3 \cdot 10^{-6} \tau$ and $\Delta t \approx 6 \cdot 10^{-4} \tau$ around a mean value of $\Delta t \approx 2.8 \cdot 10^{-4} \tau$. In the inset of (a), a close-up of the timestep behaviour is given, which reveals the rapid reduction of $\Delta t$ at jammed states and the quick recovery afterwards. In (b), the relative distribution of the data in (a) is illustrated. It is evident that the majority of steps can be executed with a large timestep, leading to increased performance of the BD simulation. On the other hand, in the rare event of a step which would produce large errors, the timestep is decreased appropriately to values far below those that would be chosen in a fixed-timestep BD run.}
  \label{fig:LJ_bulk}
\end{figure}

In contrast, the timestepping of an adaptive BD simulation run is shown both as a timeseries and as a histogram in fig.\ \ref{fig:LJ_bulk}.
The tolerance coefficients in eq.\ \eqref{eq:tol} are thereby set to $\epsilon_\mathrm{abs} = 0.05 \sigma$ and $\epsilon_\mathrm{rel} = 0.05$ and the $\infty$-norm is used in the reduction from particle-wise to global error \eqref{eq:e}.
One can see that large stepsizes up to $\Delta t \approx 6 \cdot 10^{-4} \tau$ occur without the error exceeding the tolerance threshold.
The majority of steps can be executed with a timestep larger than the value $\Delta t_\mathrm{fix} = 10^{-4} \tau$.
On the other hand, the algorithm is able to detect moves that would cause large errors where it decreases $\Delta t$ appropriately.
It is striking that in the shown sample, minimum timesteps as small as $\Delta t = 3 \cdot 10^{-6} \tau$ occur.
This is far below the stepsize of $\Delta t_\mathrm{fix} = 10^{-4} \tau$ chosen in the fixed-timestep BD run above, which indicates that although the simulation is stable for this value, there are still steps which produce substatial local errors in the particle trajectories.
For even larger values of $\Delta t_\mathrm{fix}$, it is those unresolved collision events that cause unphysical particle displacements which then cascade and crash the simulation run.
In comparison, the adaptive BD run yields a mean timestep of $\overline{\Delta t} = 3 \cdot 10^{-4} \tau$, which is larger than the heuristically chosen fixed timestep.

\subparagraph{Performance and overhead}
Per step, due to an additional evaluation via the Heun method \eqref{eq:embedded_Heun-Euler_Heun}, twice the computational effort is needed to calculate the deterministic forces compared to a single evaluation of the Euler-Maruyama step \eqref{eq:embedded_Heun-Euler_Euler}.
This procedure alone has the benefit of increased accuracy though, and it hence makes larger timesteps feasible.

The computational overhead due to adaptivity with RSwM3 comes mainly from storing random increments on both stacks and applying this information in the construction of new random forces.
Therefore, potential for further optimization lies in a cache-friendly organization of the stacks in memory as well as in circumventing superfluous access and copy instructions altogether.
The latter considerations suggest that avoiding rejections and more importantly avoiding re-rejections is crucial for good performance and reasonable memory consumption of the algorithm.
As already noted, in practice this can be accomplished with a small value of $q_\mathrm{min}$ that allows for a rapid reduction of $\Delta t$ in the case of unfortunate random events and a conservative value of $q_\mathrm{max}$ to avoid too large timestepping after moves with fortunate random increments.
In our implementation, the cost of RSwM routines is estimated to lie below $10\%$ of the total runtime in common situations.

\subsection{Non-equilibrium -- formation of colloidal films}
While the benefits of adaptive BD are already significant in equilibrium, its real advantages over conventional BD become particularly clear in non-equilibrium situations.
Due to the rich phenomenology -- which still lacks a thorough understanding -- and the important practical applications, the dynamics of colloidal suspensions near substrates and interfaces has been the center of attention in many recent works \cite{vanderKooijWatchingPaintDry2015,MakepeaceStratificationBinaryColloidal2017,FortiniStratificationSizeSegregation2017,FortiniDynamicStratificationDrying2016,TruemanAutostratificationDryingColloidal2012,SchulzCriticalQuantitativeReview2018,HeDynamicalDensityFunctional2021}.
Nevertheless, the simulation of time-dependent interfacial processes is far from straightforward and especially for common BD, stability issues are expected with increasing packing fraction.

In the following, we apply adaptive BD to systems of Lennard-Jones particles and simulate evaporation of the implicit solvent.
This is done by introducing an external potential that models the fixed substrate surface as well as a moving air-solvent interface.
As in Ref.\ \onlinecite{HeDynamicalDensityFunctional2021}, we set
\begin{equation}
  \label{eq:drying-Vext}
  V^{(i)}_\mathrm{ext}(z, t) = B \left( \mathrm{e}^{- \kappa (z - \sigma^{(i)} / 2)} + \mathrm{e}^{\kappa (z + \sigma^{(i)} / 2 - L(t))} \right)
\end{equation}
to only vary in the $z$-direction and assume periodic boundary conditions in the remaining two directions.
The value of $\kappa$ controls the softness of the substrate and the air-solvent interface while $B$ sets their strength.
We distinguish between the different particle sizes $\sigma^{(i)}$ to account for the offset of the particle centers where the external potential is evaluated.
The position $L(t)$ of the air-solvent interface is time-dependent and follows a linear motion $L(t) = L_0 - v t$ with initial position $L_0$ and constant velocity $v$.

In the following, systems in a box which is elongated in $z$-direction are considered.
To ensure dominating non-equilibrium effects, values for the air-solvent interface velocities are chosen which yield large Péclet numbers $\Pe^{(i)} = L_0 v / D^{(i)} \gg 1$ where $D^{(i)} = k_B T / \gamma^{(i)}$ is the Einstein-Smoluchowski diffusion coefficient and $\gamma^{(i)} \propto \sigma^{(i)}$ due to Stokes.

When attempting molecular simulations of such systems in a conventional approach, one is faced with a non-trivial choice of the timestep length $\Delta t_\mathrm{fix}$ since it has to be large enough to be efficient in the dilute phase but also small enough to capture the motion of the dense final state of the system.
A cumbersome solution would be a subdivision of the simulation into subsequent time intervals, thereby choosing timesteps that suit the density of the current state.
This method is beset by problems as the density profile becomes inhomogeneous and is not known a priori.

We show that by employing the adaptive BD method of Sec.\ \ref{sec:Application_to_BD}, these issues become non-existent.
Concerning the physical results of the simulation runs, the automatically chosen timestep is indeed closely connected to the increasing packing fraction as well as to the structural properties of the respective colloidal system, as will be discussed below.
Similar test runs as the ones shown in the following but carried out with constant timestepping and a naive choice of $\Delta t_\mathrm{fix}$ frequently lead to instabilities in the high-density regimes, ultimately resulting in unphysical trajectories or crashes of the program.
Due to the possiblity of stable and accurate simulations of closely packed phases with adaptive BD, we focus on the investigation of the final conformation of the colloidal suspension.

\subsubsection{Single species, moderate driving}
Firstly, a single species Lennard-Jones system is studied and the box size is chosen as $8 \times 8 \times 50 \sigma^3$.
The Lennard-Jones particles are initialized on a simple cubic lattice with lattice constant $2 \sigma$ and the velocity of the air-solvent interface is set to $v = 1 \sigma / \tau$.
We set $\epsilon_\mathrm{abs} = 0.01 \sigma$ to accomodate for smaller particle displacements in the dense phase and relax the relative tolerance to $\epsilon_\mathrm{rel} = 0.1$.

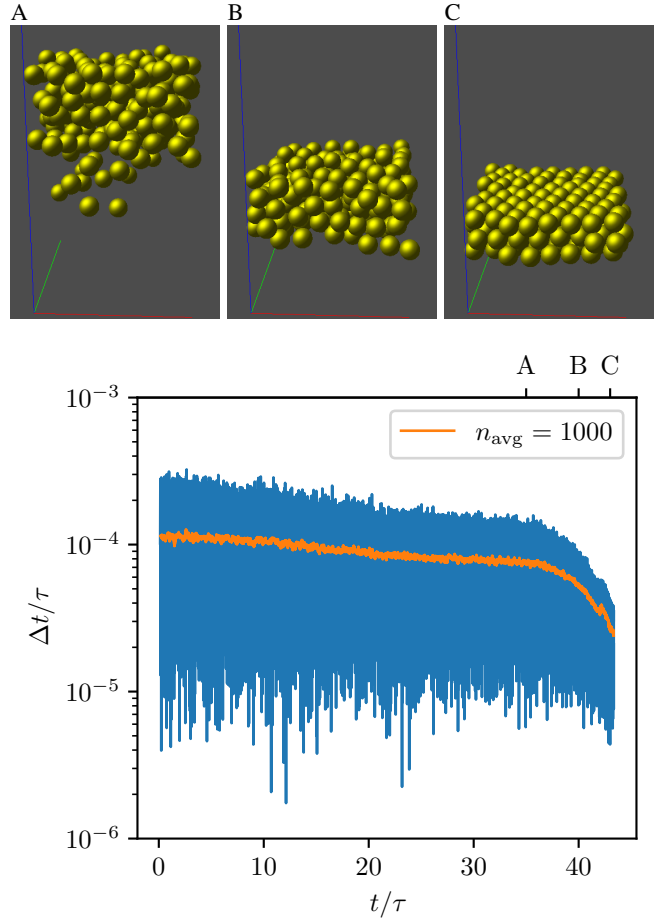
\begin{figure}[htbp]
  \centering
\begingroup%
  \makeatletter%
  \providecommand\color[2][]{%
    \errmessage{(Inkscape) Color is used for the text in Inkscape, but the package 'color.sty' is not loaded}%
    \renewcommand\color[2][]{}%
  }%
  \providecommand\transparent[1]{%
    \errmessage{(Inkscape) Transparency is used (non-zero) for the text in Inkscape, but the package 'transparent.sty' is not loaded}%
    \renewcommand\transparent[1]{}%
  }%
  \providecommand\rotatebox[2]{#2}%
  \newcommand*\fsize{\dimexpr\f@size pt\relax}%
  \newcommand*\lineheight[1]{\fontsize{\fsize}{#1\fsize}\selectfont}%
  \ifx\svgwidth\undefined%
    \setlength{\unitlength}{242.25001144bp}%
    \ifx\svgscale\undefined%
      \relax%
    \else%
      \setlength{\unitlength}{\unitlength * \real{\svgscale}}%
    \fi%
  \else%
    \setlength{\unitlength}{\svgwidth}%
  \fi%
  \global\let\svgwidth\undefined%
  \global\let\svgscale\undefined%
  \makeatother%
  \begin{picture}(1,0.53979876)%
    \lineheight{1}%
    \setlength\tabcolsep{0pt}%
    \put(0,0.48817337){\makebox(0,0)[lt]{\lineheight{1.25}\smash{\begin{tabular}[t]{l}A\end{tabular}}}}%
    \put(0.33746128,0.48817337){\makebox(0,0)[lt]{\lineheight{1.25}\smash{\begin{tabular}[t]{l}B\end{tabular}}}}%
    \put(0.67492257,0.48817337){\makebox(0,0)[lt]{\lineheight{1.25}\smash{\begin{tabular}[t]{l}C\end{tabular}}}}%
    \put(0,0){\includegraphics[width=\unitlength,page=1]{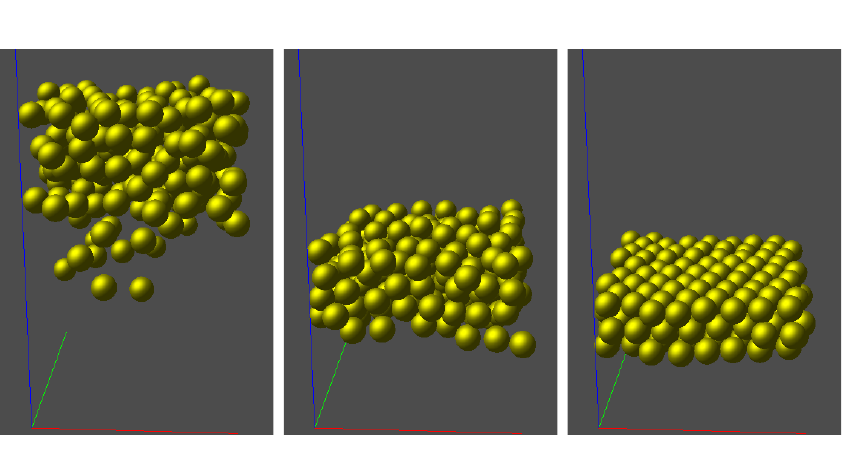}}%
  \end{picture}%
\endgroup%

  \hspace{0px}
  \includegraphics{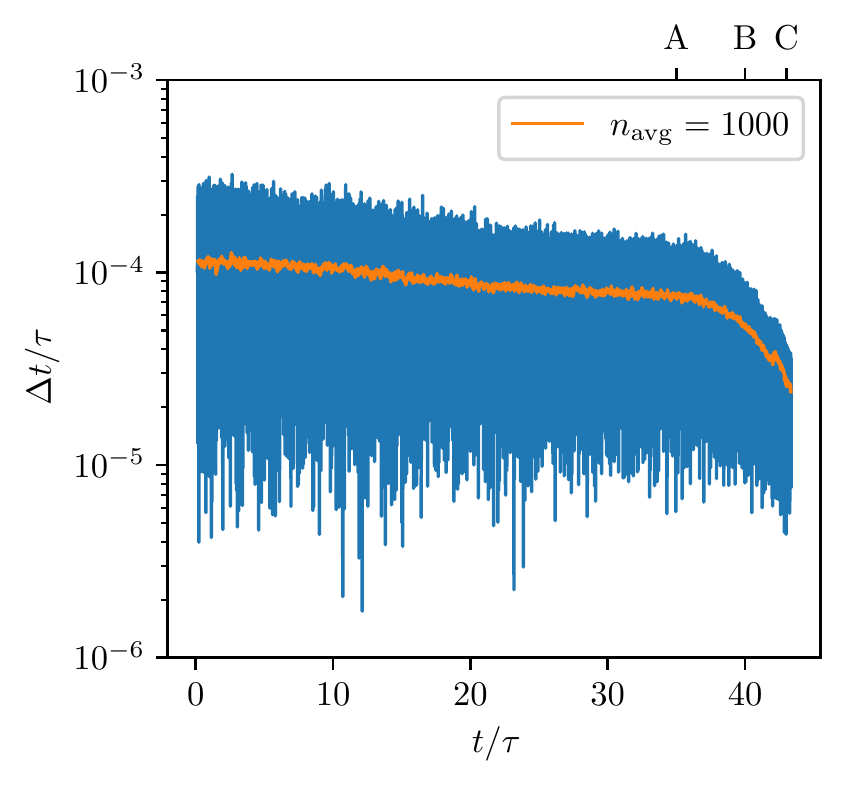}
  \caption{The timestepping evolution in a system with moving interface modeled via eq.\ \eqref{eq:drying-Vext} with parameters $B = 7000 \epsilon$, $\kappa = 1 / \sigma$ (adopted from Ref.\ \onlinecite{HeDynamicalDensityFunctional2021}), and $v = 1 \sigma / \tau$ is shown. We depict both the actual timeseries whose envelope visualizes the maximum and minimum values of $\Delta t$ as well as a moving average over the surrounding $n_\mathrm{avg}$ points to uncover the mean chosen timestep. As the density increases and the propagation of the overdamped Langevin equation becomes more difficult, the timestep is systematically decreased. This is an automatic process that needs no user input and that can even handle the freezing transition which occurs at the end of the simulation. To illustrate this process, typical snapshots of the system are given at different timepoints which are marked in the timestepping plot as A, B, and C and correspond to a dilute, a prefreezing and a crystallized state. Especially in the transition from B to C, frequent particle collisions occur which are resolved carefully by the adaptive BD method.}
  \label{fig:LJ_movingInterface}
\end{figure}

As one can see in fig.\ \ref{fig:LJ_movingInterface}, the timestep is automatically adjusted as a reaction to the increasing density.
In the course of the simulation run, the average number density increases from approximately $0.07 \sigma^{-3}$ to $1.4 \sigma^{-3}$, although the particles first accumulate near the air-solvent interface.
Astonishingly, even the freezing transition at the end of the simulation run can be captured effortlessly.
This illustrates the influence of collective order on the chosen timestep.
With rising density of the Lennard-Jones fluid, the timestep decreases on average due to the shorter mean free path of the particles and more frequent collisions.
At this stage, $\Delta t$ varies significantly and very small timesteps maneuver the system safely through jammed states of the disordered fluid.
In the process of crystallization, the timestep decreases rapidly to accomodate for the reduced free path of the particles before it shortly relaxes to a plateau when crystal order is achieved.
Additionally, the variance of $\Delta t$ decreases and, contrary to the behaviour in the liquid phase, fewer jammed states can be observed.
This is due to the crystal order of the solid phase, which prevents frequent close encounters of particles and hence alleviates the need for a rapid reduction of $\Delta t$.

\subsubsection{Single species, strong driving}
\begin{figure}[htbp]
  \centering
  \includegraphics{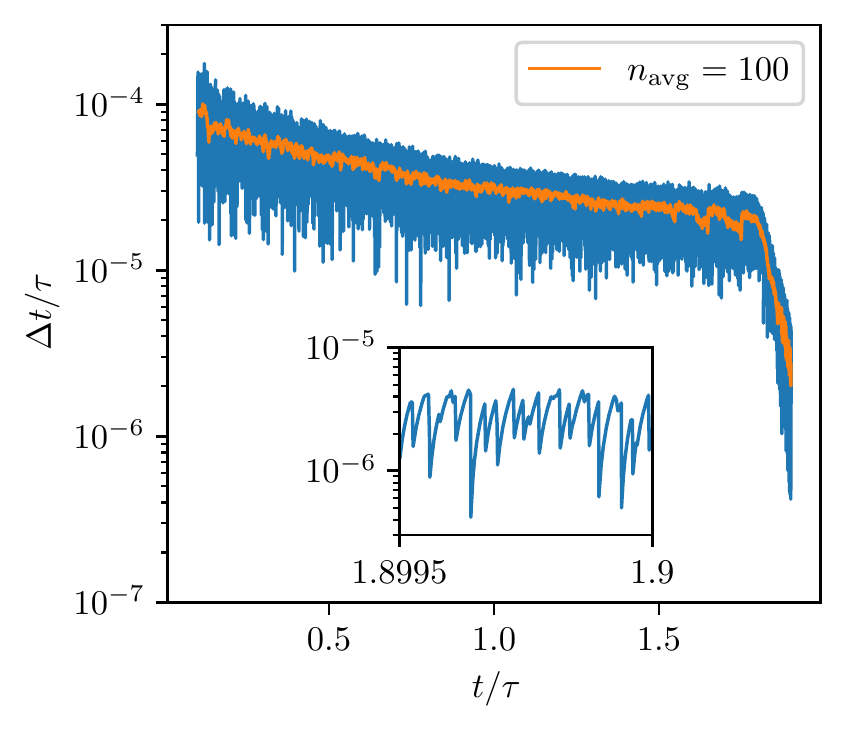}
  \caption{Time evolution and moving average of the timestep $\Delta t$ for the air-solvent interface velocity being increased to $v = 50 \sigma / \tau$ in a larger system of box size $10 \times 10 \times 100 \sigma^3$. In this case, defects are induced during the crystallization process which prevent a perfect crystal order of the final particle configuration. Jammed states still occur frequently, and the timestep has to accomodate rapidly to resolve particle displacements correctly. This is depicted in the inset, which shows that sudden jumps to small values of $\Delta t$ still occur in the high-density regime, indicating error-prone force evaluations due to prevailing defects in the crystal structure.}
  \label{fig:LJ_movingInterface_fast}
\end{figure}

If the evaporation rate is increased via a faster moving air-solvent interface, the final structure of the colloidal suspension is altered.
Particularly, with rising air-solvent velocity $v$, a perfect crystallization process is hindered and defects in the crystal structure occur.
This is reflected in the timestepping evolution, as jammed states still happen frequently in the dense regime due to misaligned particles.
Therefore, unlike in the previous case of no defects, sudden jumps to very small timesteps can still be observed as depicted in fig.\ \ref{fig:LJ_movingInterface_fast}.

If the velocity of the air-solvent interface is increased even further, amorphous states can be reached, where no crystal order prevails.
Nevertheless, our method is still able to resolve the particle trajectories of those quenched particle configurations so that the simulation remains both stable and accurate even in such demanding circumstances.

\subsubsection{Binary mixture}
Returning to stratification phenomena, we consider mixtures of particles differing in size.
Depending on the Péclet numbers of the big (subscript $b$) and small (subscript $s$) particle species and their absolute value (i.e.\ if $\Pe \ll 1$ or $\Pe \gg 1$), different structures of the final phase can emerge, ranging from ``small-on-top'' or ``big-on-top'' layering to more complicated conformations \cite{HeDynamicalDensityFunctional2021}.
For large Péclet numbers $1 \ll \Pe_s < \Pe_b$, studies of \citeauthor{FortiniDynamicStratificationDrying2016} \cite{FortiniDynamicStratificationDrying2016} have shown the formation of a ``small-on-top'' structure, i.e.\ the accumulation of the small particle species near the moving interface.
Additionally, in the immediate vicinity of the air-solvent boundary, a thin layer of big particles remains trapped due to their low mobility.

In the following, a binary mixture of Lennard-Jones particles with diameters $\sigma_b = \sigma$ and $\sigma_s = 0.5 \sigma$ is simulated in a system of size $10 \times 10 \times 100 \sigma^3$ and the velocity of the air-solvent interface is set to $v = 1 \sigma / \tau$ as before.
We initialize $N_b = 768$ big and $N_s = 4145$ small particles uniformly in the simulation domain and particularly focus in our analysis on the structure of the final dense phase.

As the simulation advances in time, the observations of \citeauthor{FortiniDynamicStratificationDrying2016} \cite{FortiniDynamicStratificationDrying2016} can be verified.
A thin layer of big particles at the air-solvent interface followed by a broad accumulation of small particles emerges.
The timestepping evolution shows similar behaviour to the single species case shown in fig.\ \ref{fig:LJ_movingInterface}.
On average, the value of $\Delta t$ decreases and throughout the simulation, jumps to low values occur repeatedly when interparticle collisions have to be resolved accurately.

As the system approaches the dense regime, the finalizing particle distribution of the dried colloidal suspension can be investigated.
One can see that a thin layer of big particles develops in close proximity to the substrate, similar to the one forming throughout the simulation at the air-solvent interface.
This process can again be explained by the lower mobility of the big particles compared to the small ones, which prevents a uniform diffusion away from the substrate.

Moreover, as the packing fraction increases further, the structure of the interfacial layers of the trapped big species changes.
While only a single peak is visible at first, a second peak develops in the last stages of the evaporation process.
This phenomenon occurs both at the substrate as well as at the air-solvent interface, although its appearance happens earlier and more pronounced at the former.

\begin{figure}[htbp]
  \centering
  \includegraphics{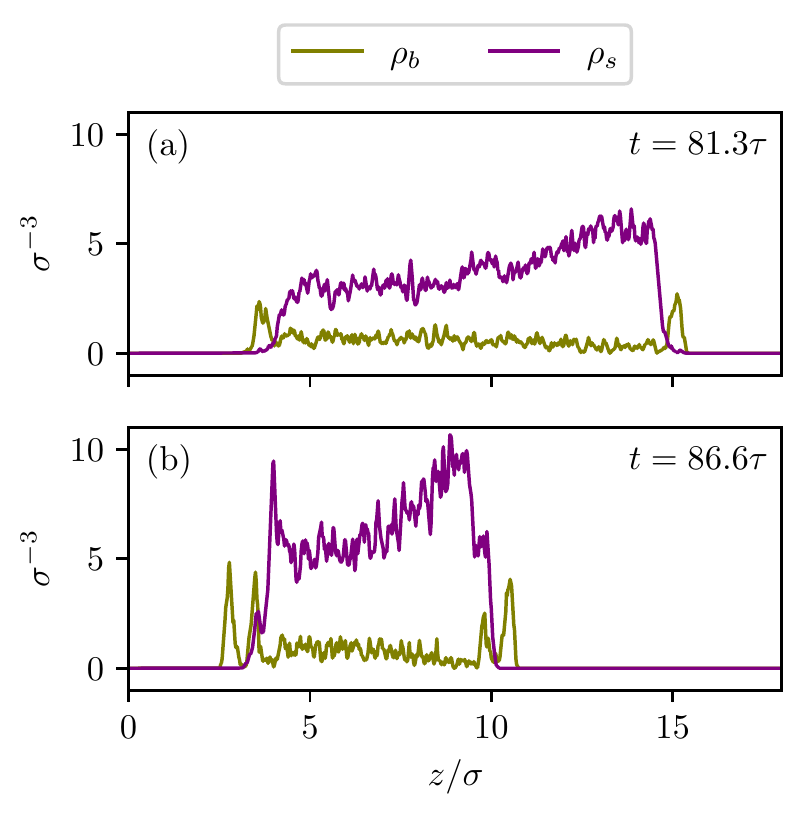}
  \caption{The density profiles $\rho_b(z)$ and $\rho_s(z)$ of the big and small particle species in a stratifying colloidal suspension of a binary Lennard-Jones mixture with particle diameters $\sigma_b = \sigma$ and $\sigma_s = 0.5 \sigma$ is shown at two timepoints of the simulation. In (a), single layers of the big species have already emerged near the substrate and the air-solvent interface, which enclose the dominating small particles in the middle of the box. At a later time (b) when the air-solvent interface has moved further towards the substrate and the packing fraction has hence increased, a second layer of the big particles forms at both interfaces and the final concentration gradient of the small species manifests within the dried film. Crucially, the intricate details of the final conformation demand an accurate numerical treatment of the dynamics of the closely packed colloidal suspension, to which adaptive BD offers a feasible solution.}
  \label{fig:LJ_stratification}
\end{figure}

Even for this simple model mixture of colloidal particles which differ only in diameter, the final conformation after evaporation of the implicit solvent possesses an intricate structure.
Both at the substrate as well as at the top of the film, a primary and secondary layer of the big particle species build up.
Those layers enclose a broad accumulation of the small particle species, which is by no means uniform but rather develops a concentration gradient in the positive $z$-direction, outlined by peaks of the respective density profile close to the big particle layers.
The formation of the described final state is illustrated in fig.\ \ref{fig:LJ_stratification}, where the density profiles of both particle species are shown for two timepoints at the end of the simulation run.

\section{Conclusion}
\label{sec:Conclusion}
In this work, we have constructed a novel method for BD simulations by employing recently developed algorithms for the adaptive numerical solution of SDEs to the case of Brownian motion as described on the level of the overdamped Langevin equation \eqref{eq:overdamped_Langevin}.
For the evaluation of a local error estimate in each trial step, we have complemented the simple Euler-Maruyama scheme \eqref{eq:embedded_Heun-Euler_Euler} found in common BD with a higher-order Heun step \eqref{eq:embedded_Heun-Euler_Heun}.
By comparison of their discrepancy with a user-defined tolerance \eqref{eq:tol} composed of an absolute and a relative contribution, we were able to impose a criterion \eqref{eq:q} for the acceptance or rejection of the trial step and for the adaptation of $\Delta t$.
Special care was thereby required in the reduction from particle-wise errors \eqref{eq:E} to a global scalar error estimate \eqref{eq:e}.

Due to the stochastic nature of Brownian motion, the rejection and subsequent retrial of a timestep could not be done naively by redrawing uncorrelated random vectors in this case.
Instead, a sophisticated method had to be employed to ensure the validity of the statistical properties of the random process in eq.\ \eqref{eq:overdamped_Langevin} and hence to avoid biased random forces from a physical point of view.
We have illustrated that the Brownian bridge theorem \eqref{eq:Brownian_bridge} resolves this problem formally, and that RSwM \cite{RackauckasAdaptiveMethodsStochastic2017} provides a feasible implementation method based on this theorem.
Hence, we specialized RSwM to the case of Brownian motion in Sec.\ \ref{subsec:RSwM_BD} and constructed the fully adaptive BD scheme outlined in alg.\ \ref{alg:RSwM3_BD}.
A correction to the original algorithm of Ref.\ \onlinecite{RackauckasAdaptiveMethodsStochastic2017} is given in Appendix \ref{appendix:RSwM3_correctness} and a generalization of adaptive BD to non-overdamped Langevin dynamics is outlined in Appendix \ref{appendix:general_Langevin}.

To test our framework, we applied the adaptive BD method to both equilibrium and non-equilibrium Lennard-Jones systems focusing on the analysis of individual trajectories.
Even in the standard case of a phase-separating bulk fluid, we could verify that the use of RSwM induced no significant computational overhead and that a performance gain could be achieved compared to the fixed-timestep Euler-Maruyama method.
This is complemented by practical convenience, since $\Delta t$ needs not be chosen a priori.

The real advantages of adaptive BD become clear in more demanding situations where a fixed timestep would ultimately lead to inaccuracy and instability of the simulation without manual intervention.
We have shown this with non-equilibrium systems of drying colloidal suspensions and have modeled the rapid evaporation of the implicit solvent by a moving interface as in Ref.\ \onlinecite{HeDynamicalDensityFunctional2021}.
With our method, we achieved efficient timestepping and unconditional stability of the simulation even in the final stages where the packing fraction increases rapidly and the final structural configuration of the dried film develops.
Particularly, in a single-species Lennard-Jones system, the freezing transition could be captured effortlessly even if crystal defects remained as a result of strong external driving.
In the more elaborate case of a binary mixture of different-size particles, the intricate structure of the colloidal film could be resolved accurately.
Here, we reported the development of a dual layer of the big particle species both at the substrate as well as at the top interface, while a concentration gradient of the small particle species could be observed in-between.
This shows that the method can be used to predict the arrangement of stratified films efficiently and with great detail, which could be helpful in the determination of their macroscopic properties.

We expect similar benefits in all sorts of situations where fundamental changes within the system call for more accurate or more relaxed numerical timestepping.
For instance, this could be the case in sheared systems \cite{LauratiTransientDynamicsDense2012,JahreisShearinducedDeconfinementHard2020}, for sedimenting \cite{ArcherInterplaySedimentationPhase2011,RoyallNonequilibriumSedimentationColloids2007} or cluster-forming \cite{BleibelShockWavesCapillary2011} colloidal suspensions, in microrheological simulations \cite{CarpenMicrorheologyColloidalDispersions2005}, for phase transitions and especially the glass transition \cite{LowenBrownianDynamicsKinetic1991}, in the formation of crystals and quasicrystals \cite{ArcherSoftcoreParticlesFreezing2015,ArcherQuasicrystallineOrderCrystalLiquid2013}, in active matter \cite{VolpeSimulationActiveBrownian2014,FarageEffectiveInteractionsActive2015,PaliwalNonequilibriumSurfaceTension2017,PaliwalChemicalPotentialActive2018}, possibly including active crystallization \cite{OmarPhaseDiagramActive2021,TurciPhaseSeparationMultibody2021}, as well as for responsive colloids \cite{BaulStructureDynamicsResponsive2021}.
In any respect, adaptive BD can help in a more accurate measurement of observables, since the evaluation of the average \eqref{eq:Sampling_timeaverage_general} is not hampered by erroneous particle configurations, which might occur in conventional BD.
Especially for quantities with a one-sided bias such as absolute or square values of particle forces and velocities, the abundance of outliers in the set of samples used in eq.\ \eqref{eq:Sampling_timeaverage_general} is essential to yield valid results.

We finally illustrate possible improvements of the adaptive BD method itself, where several aspects come to mind.
Firstly, the choices of the parameters $\alpha$ in eq.\ \eqref{eq:q} as well as $q_\mathrm{min}$ and $q_\mathrm{max}$ in eq.\ \eqref{eq:q_bounded} were made mostly heuristically.
We note that in Ref.\ \onlinecite{RackauckasAdaptiveMethodsStochastic2017}, the influence of those parameters on the performance of the algorithm has been investigated, but emphasize that results could vary for our high-dimensional setting of Brownian motion.

Secondly, recently developed adaptation methods for the stepsize which employ control theory could be helpful.
Thereby, $\Delta t$ is not merely scaled after each trial step by a locally determined factor $q$ as evaluated in eq.\ \eqref{eq:q}.
Instead, one constructs a proportional-integral-derivative (PID) controller for the selection of new timesteps.
This means that the adaptation is now not only influenced proportionally by the error estimate of the current step, but rather it is also determined by integral and differential contributions, i.e.\ the memory of previous stepsizes and the local change of the stepsize.
Adaptive timestepping with control theory has already been applied to ODEs \cite{GustafssonControltheoreticTechniquesStepsize1994, SoderlindAutomaticControlAdaptive2002, SoderlindDigitalFiltersAdaptive2003} and SDEs \cite{BurrageAdaptiveStepsizeBased2004}.
This method could help in our case to reduce the number of unneccessarily small steps even further.
In our present implementation, a conservatively chosen $q_\mathrm{max}$ in eq.\ \eqref{eq:q_bounded} keeps the number of rejections low because it restricts the growth of $\Delta t$ after moves with coincidentally low error.
But it also has the effect of preventing a fast relaxation of $\Delta t$ after a sudden drop, e.g.\ due to an unfortunate random event.
Then, many steps are needed for $\Delta t$ to grow back to its nominal value, because the maximum gain in each step is limited by $q_\mathrm{max}$.
With a control theory approach, a mechanism could be established which permits sudden drops to low values of $\Delta t$ but still ensures a rapid relaxation afterwards.

Finally, the adaptive BD method could be augmented to include hydrodynamic interactions.
This requires a careful treatment of the random forces, as they now incorporate the particle configuration $\vec{r}^N$.
Therefore, the noise is no longer additive, which complicates the numerical scheme necessary for a correct discretisation of the overdamped Langevin equation.
Still, instead of eqs.\ \eqref{eq:embedded_Heun-Euler_Euler} and \eqref{eq:embedded_Heun-Euler_Heun}, existing methods for SDEs with non-additive noise could be employed within the presented framework to yield an adaptive timestepping procedure for BD with hydrodynamic interactions.

In summary, to capture the dynamics of systems that undergo structural changes such as the ones shown above, and to gain further understanding of the implications regarding e.g.\ the resulting macroscopic properties, sophisticated simulation methods are needed.
Adaptive BD provides the means to treat these problems with great accuracy while still being computationally feasible and robust, which is a crucial trait from a practical standpoint.

\begin{acknowledgments}
  We thank Christopher Rackauckas, Tobias Eckert and Daniel de las Heras for useful comments.
  This work is supported by the German Research Foundation (DFG) via project number 436306241.
\end{acknowledgments}

\section*{Data availability statement}
The data that support the findings of this study are available from the corresponding author upon reasonable request.

\bibliography{bibliography.bib}

\appendix

\section{Generalization to Langevin dynamics}
\label{appendix:general_Langevin}
In the following, we transfer the concepts of Sec.\ \ref{sec:Application_to_BD} to general (non-overdamped) Langevin dynamics.
Thereby, the momenta of particles with masses $m^{(i)}$, $i = 1,\dots,N$, are explicitly considered in the Langevin equation
\begin{equation}
  \label{eq:general_Langevin}
  m^{(i)} \ddot{\vec{r}}^{(i)}(t) = \vec{F}^{(i)}(\vec{r}^N(t)) - \gamma^{(i)} \dot{\vec{r}}^{(i)}(t) + \sqrt{2 \gamma^{(i)} k_B T} \vec{R}^{(i)}(t)
\end{equation}
where the notation is that of eq.\ \eqref{eq:overdamped_Langevin}.
Depending on the choice of $\gamma^{(i)}$, two special cases can be identified.
BD is recovered in the diffusive regime for large $\gamma^{(i)}$.
On the other hand, for vanishing friction and random forces, i.e.\ $\gamma^{(i)} = 0$, deterministic Hamiltonian equations of motion are obtained.

The SDE \eqref{eq:general_Langevin} can be reformulated as a pair of first-order equations for the particle positions $\vec{r}^N$ and velocities $\vec{v}^N = \dot{\vec{r}}^N$,
\begin{subequations}
  \label{eq:general_Langevin_rv}
  \begin{align}
    \begin{split}
      m^{(i)} \dot{\vec{v}}^{(i)}(t) &= \vec{F}^{(i)}(\vec{r}^N(t)) - \gamma^{(i)} \vec{v}^{(i)}(t)\\
      &\qquad + \sqrt{2 \gamma^{(i)} k_B T} \vec{R}^{(i)}(t),
    \end{split}\\
    \dot{\vec{r}}^{(i)}(t) &= \vec{v}(t).
  \end{align}
\end{subequations}
From a numerical perspective, the treatment of eqs.\ \eqref{eq:general_Langevin_rv} differs significantly from that of the overdamped Langevin equation \eqref{eq:overdamped_Langevin}, since the second-order nature makes more involved timestepping procedures for $\vec{r}^N$ and $\vec{v}^N$ possible.

This can be illustrated easily when eqs.\ \eqref{eq:general_Langevin_rv} are considered in the Hamiltonian case.
Then, via the separation of $\vec{r}^N$ and $\vec{v}^N$, simple semi-implicit methods are readily available, such as the Euler-Cromer or the well-known velocity Verlet method which is commonly used in MD \cite{VerletComputerExperimentsClassical1967,SwopeComputerSimulationMethod1982}.
From a fundamental point of view, the use of symplectic algorithms is necessary for Hamiltonian systems to ensure the correct description of physical conservation laws and hence to achieve numerical stability for long-time behaviour.
While the above mentioned integrators possess this property, most explicit algorithms like the forward Euler and classic Runge-Kutta method fail in this regard.

When going from Hamiltonian to Langevin dynamics, one is faced with an appropriate generalization of the familiar deterministic integration schemes to include dissipative and random forces.
This is addressed by so-called quasi-symplectic methods, which are integrators for SDEs of type \eqref{eq:general_Langevin_rv} that degenerate to symplectic ones when both the noise and the friction term vanish \cite{MilsteinQuasisymplecticMethodsLangevintype2003}.
Still, it is not straightforward to construct schemes that are suitable for arbitrary $\gamma^{(i)}$, and different strategies have been used in Langevin-based molecular simulations \cite{BurrageNumericalMethodsSecond2007,vanGunsterenAlgorithmsBrownianDynamics1982,BrungerStochasticBoundaryConditions1984,SkeelImpulseIntegratorLangevin2002,GogaEfficientAlgorithmsLangevin2012,LeimkuhlerMolecularDynamics2015}.

To incorporate adaptive timestepping, one must again choose an integrator pair of different order to calculate an error estimation per step.
The above considerations suggest however, that at least for the higher-order method, a quasi-symplectic scheme is advisable for optimal efficiency and accuracy as it is expected to perform better than a naive application of the embedded Heun-Euler method \eqref{eq:embedded_Heun-Euler_Euler} and \eqref{eq:embedded_Heun-Euler_Heun} to general Langevin dynamics.
Nevertheless, the latter could still be an adequate option when friction dominates.

Then, with the formalism described in Sec.\ \ref{sec:Application_to_BD}, the acceptance or rejection of trial steps and the use of RSwM is straightforward, so that adaptive algorithms involving the Langevin equation \eqref{eq:general_Langevin} could be a feasible means for the simulation of dissipative particle dynamics in the future.

\section{Correctness of RSwM3 rejection branch}
\label{appendix:RSwM3_correctness}
Although the effect is likely subtle or even unmeasurable in most situations, in the original RSwM3 algorithm as described in Ref.\ \onlinecite{RackauckasAdaptiveMethodsStochastic2017} and formerly implemented in DifferentialEquations.jl \cite{DifferentialEquations.jl-2017}, the $q < 1$ branch was handled incorrectly.
We list a pseudocode version thereof in alg.\ \ref{alg:RSwM3_old_rejection}, where a specialization to our case of BD is already performed for ease of comparison with alg.\ \ref{alg:RSwM3_BD}.

\begin{figure}[htbp]
  \centering
  \input{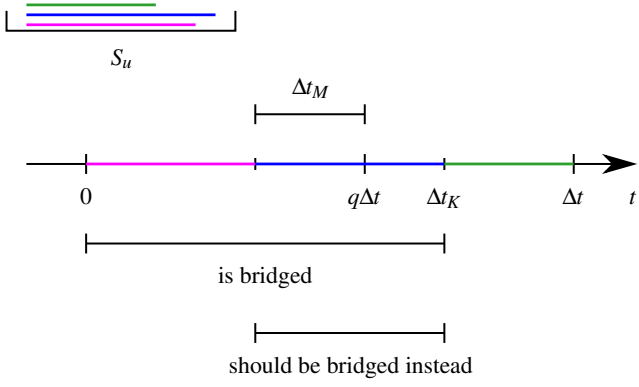}
  \caption{A scenario is shown for which the original implementation of the RSwM3 rejection branch fails. For clarity, we only display the time intervals which reside on the stack $S_u$ without their corresponding random vectors. In the presented case, the wrong interval (``is bridged'') is selected in the original RSwM3 implementation to which the Brownian bridge theorem \eqref{eq:Brownian_bridge} is applied. This is fixed in alg.\ \ref{alg:RSwM3_BD}, where the correct element (``should be bridged'') is considered via the calculation of $\Delta t_M$.}
  \label{fig:rswm3_correctness}
\end{figure}

To see why alg.\ \ref{alg:RSwM3_old_rejection} fails in some cases, consider multiple elements present on the stack $S_u$, e.g.\ as depicted in fig.\ \ref{fig:rswm3_correctness} where $S_u$ contains three elements.
Then in lines 2--11 of alg.\ \ref{alg:RSwM3_old_rejection}, elements are transferred successively from $S_u$ back to $S_f$ and their time intervals are accumulated in $\Delta t_s$ as long as the remainder $\Delta t - \Delta t_s$ is still larger than the next goal timestep $q \Delta t$ (this is only the case for the green element in fig.\ \ref{fig:rswm3_correctness}).
After that, the Brownian bridge theorem is applied to $(\Delta t_K, \vec{R}_K) = (\Delta t - \Delta t_s, \vec{R} - \vec{R}_s)$.
Therefore, the interpolation includes all the remaining elements of the stack $S_u$, i.e.\ the blue and violet one in fig.\ \ref{fig:rswm3_correctness}, which implies that intermediate values of the Brownian path are not considered if $S_u$ holds two or more elements at this point.

To yield a valid interpolation in this situation, the Brownian bridge must instead only be applied to the immediately following single element at the top of $S_u$ (the blue one in fig.\ \ref{fig:rswm3_correctness}).
In alg.\ \ref{alg:RSwM3_BD}, lines 4--19 therefore replace the logic of alg.\ \ref{alg:RSwM3_old_rejection}, which fixes the above issue by always considering only the single element that crosses $q \Delta t$ in the application of eq.\ \eqref{eq:Brownian_bridge}.
This element is then interpolated at $q \Delta t$, for which the variable $\Delta t_M$ is introduced to yield the corresponding fraction $q_M$ of the element.
Note that $S_u$ always contains at least one element when entering the rejection branch (assuming a proper initialization of the simulation, cf.\ Appendix \ref{appendix:RSwM3_BD}).

\begin{algorithm}[H]
  \caption{Former $q < 1$ branch of RSwM3 where the bridge theorem might be applied to a wrong interval}
  \label{alg:RSwM3_old_rejection}
  \begin{algorithmic}[1]
    \State $\Delta t_s \gets 0$, $\vec{R}_s \gets 0$
    \While{$S_u$ not empty}
      \State Pop top of $S_u$ as $(\Delta t_u, \vec{R}_u)$ 
      \If{$\Delta t_s + \Delta t_u < (1 - q) \Delta t$}
        \State $\Delta t_s \gets \Delta t_s + \Delta t_u$, $\vec{R}_s \gets \vec{R}_s + \vec{R}_u$
        \State Push $(\Delta t_u, \vec{R}_u)$ onto $S_f$
      \Else
        \State Push $(\Delta t_u, \vec{R}_u)$ back onto $S_u$
        \State \textbf{break}
      \EndIf
    \EndWhile
    \State $\Delta t_K \gets \Delta t - \Delta t_s$, $\vec{R}_K \gets \vec{R} - \vec{R}_s$
    \State $q_K \gets q \Delta t / \Delta t_K$
    \State $\vec{R}_\mathrm{bridge} \sim \mathcal{N}(q_K \vec{R}_K, (1 - q_K) q_K \Delta t_K)$
    \State Push $((1 - q_K) \Delta t_K, \vec{R}_K - \vec{R}_\mathrm{bridge})$ onto $S_f$
    \State $\Delta t \gets q \Delta t$, $\vec{R} \gets \vec{R}_\mathrm{bridge}$
  \end{algorithmic}
\end{algorithm}

\section{Implementation of embedded Heun-Euler trial step with RSwM3}
\label{appendix:RSwM3_BD}

The embedded Heun-Euler trial step with RSwM3 is the integral part of the adaptive BD method, for which we provide a pseudocode implementation in the following and explain its technical details as well as its setting in a full simulation framework.
At the beginning of the simulation, $S_f$ must be empty and a finite positive value of the first timestep $\Delta t$ is chosen heuristically (this choice is irrelevant, however, since the algorithm immediately adapts $\Delta t$ to acceptable values).
Gaussian random increments are drawn for the initial trial step via $\vec{R} \sim \mathcal{N}(0, \Delta t)$ which are pushed onto the stack $S_u$.
This completes the initialization of the simulation run and a loop over trial steps can be started (until a certain simulation time or number of steps is reached).

Then, in each trial step, which is listed in alg.\ \ref{alg:RSwM3_BD}, the Euler and Heun approximations $\bar{\vec{r}}_{k+1}$ and $\vec{r}_{k+1}$ are evaluated via eqs.\ \eqref{eq:embedded_Heun-Euler_Euler} and \eqref{eq:embedded_Heun-Euler_Heun} and the adaptation factor $q$ is calculated via eqs.\ \eqref{eq:q} and \eqref{eq:q_bounded}.
This requires an error estimate \eqref{eq:e} which is obtained from the two approximations via eqs.\ \eqref{eq:E} and \eqref{eq:tol}.

If $q < 1$, the proposed step is rejected and a retrial must be performed.
For a detailed description of the rejection branch (lines 4--19), we refer to Appendix \ref{appendix:RSwM3_correctness}, where our changes to the original RSwM3 $q < 1$ case are illustrated as well.
If a step is accepted, i.e.\ $q > 1$, the particle positions and physical time of the system are updated accordingly (line 21) before resetting random increments $\vec{R}$ and stack $S_u$ (line 22).
Then, in lines 23--42, new random increments are constructed for the next trial step.

For this, elements on $S_f$ stemming from previously rejected steps have to be accounted for as long as parts of them lie within the goal timestep $\Delta t$.
Therefore, the elements are transferred successively from $S_f$ to $S_u$ and their time intervals and random increments are accumulated in $\Delta t_s$ and $\vec{R}$.
If at some point $\Delta t_s$ exceeds $\Delta t$, the corresponding element is interpolated at $\Delta t$ via the Brownian bridge theorem \eqref{eq:Brownian_bridge} in lines 29--34 and one proceeds with the next trial step.
Conversely, if all elements of $S_f$ were popped and a gap $\Delta t_\mathrm{gap}$ remains between $\Delta t_s$ and the goal timestep $\Delta t$, new Gaussian random increments $\vec{R}_\mathrm{gap} \sim \mathcal{N}(0, \Delta t_\mathrm{gap})$ are drawn, added to $\vec{R}$, and pushed onto $S_u$ in lines 39--41 before attempting the next trial step.
Note that the values of $\vec{R}$ persist across trial steps and are only reset after accepted moves.

Throughout the algorithm, bookkeeping of random increments and corresponding time intervals is established by pushing those elements that are involved in the current random increment $\vec{R}$ onto $S_u$, and by storing elements that become relevant beyond the current step on $S_f$.
This procedure is especially important after bridging an element at the goal timestep, where the two resulting parts are pushed onto $S_u$ and $S_f$ respectively (lines 14, 15 and 31, 32).
Consequently, no drawn random increments are ever lost.

\begin{algorithm}[H]
  \caption{Embedded Heun-Euler trial step with RSwM3}
  \label{alg:RSwM3_BD}
  \begin{algorithmic}[1]
    \State Calculate $\bar{\vec{r}}_{k + 1}$ and $\vec{r}_{k + 1}$ via eqs.\ \eqref{eq:embedded_Heun-Euler_Euler} and \eqref{eq:embedded_Heun-Euler_Heun}
    \State Calculate $q$ via eqs.\ \eqref{eq:E}, \eqref{eq:tol}, \eqref{eq:e}, \eqref{eq:q} and \eqref{eq:q_bounded}
    \If{$q < 1$}  \Comment{cf.\ fig.\ \ref{fig:rejection}, Appendix \ref{appendix:RSwM3_correctness}}
      \State $\Delta t_s \gets 0$, $\vec{R}_s \gets 0$
      \While{$S_u$ not empty}
        \State Pop top of $S_u$ as $(\Delta t_u, \vec{R}_u)$
        \State $\Delta t_s \gets \Delta t_s + \Delta t_u$, $\vec{R}_s \gets \vec{R}_s + \vec{R}_u$
        \If{$\Delta t_s < (1 - q) \Delta t$}
          \State Push $(\Delta t_u, \vec{R}_u)$ onto $S_f$
        \Else
          \State $\Delta t_M \gets \Delta t_s - (1 - q) \Delta t$
          \State $q_M \gets \Delta t_M / \Delta t_u$
          \State $\vec{R}_\mathrm{bridge} \sim \mathcal{N}(q_M \vec{R}_u, (1 - q_M) q_M \Delta t_u)$
          \State Push $((1 - q_M) \Delta t_u, \vec{R}_u - \vec{R}_\mathrm{bridge})$ onto $S_f$
          \State Push $(q_M \Delta t_u, \vec{R}_\mathrm{bridge})$ onto $S_u$
          \State \textbf{break}
        \EndIf
      \EndWhile
      \State $\Delta t \gets q \Delta t$, $\vec{R} \gets \vec{R} - \vec{R}_s + \vec{R}_\mathrm{bridge}$
    \Else
      \State Do step: $t \gets t + \Delta t$, $\vec{r}_{k} \gets \vec{r}_{k + 1}$, $\Delta t \gets q \Delta t$
      \State Empty $S_u$, $\Delta t_s \gets 0$, $\vec{R} \gets 0$
      \While{$S_f$ not empty}
        \State Pop top of $S_f$ as $(\Delta t_f, \vec{R}_f)$ 
        \If{$\Delta t_s + \Delta t_f < \Delta t$}
          \State $\Delta t_s \gets \Delta t_s + \Delta t_f$, $\vec{R} \gets \vec{R} + \vec{R}_f$
          \State Push $(\Delta t_f, \vec{R}_f)$ onto $S_u$
        \Else \Comment{cf.\ fig.\ \ref{fig:setup_next_step_with_bridge}}
          \State $q_M \gets (\Delta t - \Delta t_s) / \Delta t_f$
          \State $\vec{R}_\mathrm{bridge} \sim \mathcal{N}(q_M \vec{R}_f, (1 - q_M) q_M \Delta t_f)$
          \State Push $((1 - q_M) \Delta t_f, \vec{R}_f - \vec{R}_\mathrm{bridge})$ onto $S_f$
          \State Push $(q_M \Delta t_f, \vec{R}_\mathrm{bridge})$ onto $S_u$
          \State $\Delta t_s \gets \Delta t_s + q_M \Delta t_f$, $\vec{R} \gets \vec{R} + \vec{R}_\mathrm{bridge}$
          \State \textbf{break}
        \EndIf
      \EndWhile
      \State $\Delta t_\mathrm{gap} \gets \Delta t - \Delta t_s$
      \If{$\Delta t_\mathrm{gap} > 0$} \Comment{cf.\ fig.\ \ref{fig:setup_next_step_with_gap}}
        \State $\vec{R}_\mathrm{gap} \sim \mathcal{N}(0, \Delta t_\mathrm{gap})$
        \State $\vec{R} \gets \vec{R} + \vec{R}_\mathrm{gap}$
        \State Push $(\Delta t_\mathrm{gap}, \vec{R}_\mathrm{gap})$ onto $S_u$
      \EndIf
    \EndIf
  \end{algorithmic}
\end{algorithm}

\end{document}